\documentclass[useAMS,usenatbib]{mn2e}
\usepackage{graphicx}
\usepackage{epstopdf}
\usepackage{lscape}
\usepackage{amsmath}
\usepackage{rotating}
\usepackage{natbib}
\usepackage{color}

\newcommand{\mnras}{MNRAS}
\newcommand{\aj}{AJ}
\newcommand{\apj}{ApJ}

\newcommand{\apjl}{ApJ}
\newcommand{\aap}{A\&A}

\newcommand{\araa}{Annual Review of Astronomy and Astrophysics}

\title[Velocity ellipsoid Gaia DR1]{The velocity ellipsoid in the Galactic disc using \emph{Gaia} DR1}
\author[Anguiano et al.]{Borja~Anguiano$^{1,2}$\thanks{E-mail: ba7t@virginia.edu}, Steven R.~Majewski$^{1}$, Kenneth C.~Freeman$^{3}$, Arik W.~Mitschang$^{2}$ \newauthor Martin~C.~Smith$^{4}$ \\
$^{1}$ Department of Astronomy, University of Virginia, Charlottesville, VA 22904-4325, USA\\
$^{2}$ Department of Physics $\&$ Astronomy, Macquarie University, Balaclava Rd, NSW 2109, Australia\\
$^{3}$ Research School of Astronomy $\&$ Astrophysics, Australian National University, Cotter Rd., Weston, ACT 2611, Australia\\
$^{4}$ Shanghai Astronomical Observatory, 80 Nandan Road, Shanghai 200030, China}

\begin{document}

\date{\today}

\pagerange{\pageref{firstpage}--\pageref{lastpage}} \pubyear{2016}

\maketitle

\label{firstpage}

\begin{abstract}

The stellar velocity ellipsoid of the solar neighbour (d $<$ 200 pc) is re-examined using intermediate-old mono-abundances stellar groups with high quality chemistry data together with parallaxes and proper motions from Gaia DR1. We find the average velocity dispersion values for the three space velocity components for the thin and thick disc of ($\sigma_{\rm U}$,$\sigma_{\rm V}$,$\sigma_{\rm W}$)$_{\rm thin}$ = (33 $\pm$ 4, 28 $\pm$ 2, 23 $\pm$ 2) and ($\sigma_{\rm U}$,$\sigma_{\rm V}$,$\sigma_{\rm W}$)$_{\rm thick}$ = (57 $\pm$ 6, 38 $\pm$ 5, 37 $\pm$ 4) km s$^{-1}$, respectively. The mean values of the ratio between the semi-axes of the velocity ellipsoid for the thin disc are found to be, $\sigma_{\rm V}/\sigma_{\rm U}$ = 0.70 $\pm$ 0.13 and $\sigma_{\rm W}/\sigma_{\rm U}$ is 0.64 $\pm$ 0.08, while for the thick disc $\sigma_{\rm V}/\sigma_{\rm U}$ = 0.67 $\pm$ 0.11 and $\sigma_{\rm W}/\sigma_{\rm U}$ is 0.66 $\pm$ 0.11. Inputting these dispersions into the linear Str\"omberg relation for the thin disc groups, we find the Sun's velocity with respect to the LSR in Galactic rotation to be V$_{\sun}$ = 13.9 $\pm$ 3.4 km s$^{-1}$. A relation is found between the vertex deviation and the chemical abundances for the thin disc, ranging from -5 to +40$^{\circ}$ as iron-abundance increases. For the thick disc we find a vertex deviation of l$_{\rm uv}$ $\sim$ -15$^{\circ}$. The tilt angle (l$_{\rm uw}$) in the $U$-$W$ plane for the thin disc groups ranges from -10 to +15$^\circ$, but there is no evident relation between l$_{\rm uw}$ and the mean abundances. However we find a weak relation for l$_{\rm uw}$ as a function of iron abundances and $\alpha$-elements for most of the groups in the thick disc, where the tilt angle decreases from -5 to -20$^\circ$ when [Fe/H] decreases and [$\alpha$/Fe] increases. The velocity anisotropy parameter is independent of the chemical group abundances and its value is nearly constant for both discs ($\beta$ $\sim$ 0.5), suggesting that the combined disc is dynamically relaxed.

\end{abstract}


\section{Introduction}

Determining the shape and orientation of the three-dimensional distribution of stellar velocities (i.e., the velocity ellipsoid) is a longstanding problem in Galactic dynamics. From epicyclic theory and for a galaxy with a flat rotation curve, the ratio of tangential to radial velocity dispersion is predicted to be 1/$\sqrt2$ and the orientation of the stellar velocity ellipsoid is tightly related to the shape and symmetry of the Galactic potential --- see,  for instance,  \citet{1991ApJ...368...79A, 1991ApJ...367L...9K,2012ApJ...746..181S} and references therein. On the other hand, non-axisymmetric structures such as bars \citep{2000AJ....119..800D,2010MNRAS.407.2122M} and spiral arms in disk galaxies \citep{2008MNRAS.383..817V} might play an important role in influencing the observed orientation of the stellar velocity ellipsoid. For example, \citet{2013ApJ...764..123S} used N-body simulations to follow the evolution of the stellar velocity ellipsoid in a galaxy that undergoes bar instability, and showed that the tilt of the stellar velocity ellipsoid is a very good indicator of the buckling that forms stellar bars in disk galaxies.

Measuring observationally a precise orientation of the velocity ellipsoid has been difficult, mainly due to the absence of reliable parallaxes and proper motions for stars outside the Solar vicinity in the pre-\emph{Gaia} era. We can calculate the tilt angles of the velocity ellipsoid using the following formula:

\begin{equation}
\rm tan(2 l_{ij}) = \frac{2\sigma_{ij}^2}{\sigma_{ii}^2 - \sigma_{jj}^2}
\label{eq:tilt}
\end{equation}

where the tilt angle (l$_{ij}$) corresponds to the angle between the $i$-axis and the major axis of the ellipse formed by projecting the three-dimensional velocity ellipsoid onto the $ij$-plane, where $i$ and $j$ are any of the stellar velocities. The traditional vertex deviation within the Galactic plane is defined as l$_{\rm uv}$. By extension we define the tilt of the velocity ellipsoid with respect to the Galactic plane as l$_{\rm uw}$. If the deviation is positive the long axis of the velocity ellipsoid is pointing into the first (l = 0 - 90$^{\circ}$) quadrant. In a stationary, axisymmetric disk galaxy, the stellar velocity ellipsoid in the Galactic mid-plane is perfectly aligned with the Galactocentric coordinate axes (e.g., \citealt{2008gady.book.....B, 2012ApJ...746..181S}).

\citet{2008MNRAS.391..793S} and \citet{2012A&A...547A..70P}, using a stellar sample from the RAVE survey covering a relatively large range in height from the plane (0.5 $<$ $z$ $<$ 1.5 kpc), found that the local velocity ellipsoid is tilted toward the Galactic Plane, with an angle of +7.3 $\pm$ 1.8$^{\circ}$. Using also radial velocities from the RAVE survey together with proper motions from the Southern Proper Motion Program (SPMP), \citet{2011ApJ...728....7C} found a tilt angle of +8.6 $\pm$ 1.8$^{\circ}$. From SDSS DR7 disc stars, \citet{2010ApJ...712..692C} found a tilt angle of +7.1 $\pm$ 1.5$^{\circ}$ while for very metal-poor halo stars they found a larger tilt of +10.3 $\pm$ 0.4$^{\circ}$ in the range of Galactic heights 1 $<$ z $<$ 2 kpc. \citet{2012ApJ...747..101M}, used a sample of $\sim$ 1200 red giants vertically distributed in a cone of 15$^{\circ}$ radius centered on the South Galactic Pole, found results in agreement with those of \citet{2008MNRAS.391..793S} and \citet{2011ApJ...728....7C}. \citet{2012ApJ...746..181S} also used SDSS DR7 data, but restricted to Stripe 82. These authors concluded that the tilt angles are consistent with a velocity ellipsoid aligned in spherical polar coordinates, however, the data suffers from larger uncertainties. Recently, \citet{2015MNRAS.452..956B}, using SDSS/SEGUE G-type dwarfs stars together with USNO-B proper motions, found for the Solar neighborhood ($\mid$$z$$\mid$ $<$ 425 pc) a tilt angle of --4.7 $\pm$ 2.0$^{\circ}$. All of these previous studies addressing the orientation of the velocity ellipsoid problem considered the Galactic disc as a single component, with the exception of \citet{2015MNRAS.452..956B} who distinguished between two sub-samples using [$\alpha$/Fe] abundances. However, it is well established that the Galactic disc has a thick component that is chemically distinguished from a thinner one, suggesting a different chemical evolution and, hence distinct disc-formation mechanisms and epochs (e.g.; \citealt{1997ApJ...477..765C,2011MNRAS.414.2893F,2015MNRAS.453.1855M}). This high-[$\alpha$/Fe] sequence, related to the thick disc, exists over a large radial and vertical range of the Galactic disc (e.g., \citealt{2014ApJ...796...38N,2015ApJ...808..132H}).

One of the main goals of the present work is to explore the orientation of the velocity ellipsoid for the thin and thick discs separately and in mono-abundances groups. Using a very local sample close to the Galactic plane, where the expected velocity ellipsoid tilt should be zero, we want to quantify possible tilt deviations related to the non-axisymmetric structures in the Milky Way disc. We also explore the ellipsoid orientation as a function of stellar iron and [$\alpha$/Fe] abundances. To understand the kinematical properties in the solar circle for the two components of the Galactic disc, we distinguish the two components chemically employing \emph{chemical labeling}, where stars are grouped in abundance space \citep{2002ARA&A..40..487F,2012MNRAS.421.1231T,2013MNRAS.428.2321M}. We also make use of parallaxes and proper motions from the first \emph{Gaia} data release \citep{2016A&A...595A...2G}. Variations of the vertex deviation and the tilt angle of the velocity ellipsoid with respect to the vertical Galactic height have been the subject of several studies \citep{2012ApJ...746..181S,2012ApJ...747..101M,2015MNRAS.452..956B}, however little progress has been made to address the orientation of the velocity ellipsoid with respect to stellar abundances. Moreover, we exploit the derived velocity dispersions to explore the relation between the Galactic rotation velocity and the stellar abundances. Recently, an intriguing statistical relation has been found between rotational velocity and [Fe/H], with opposite signs for the thin and thick disc \citep{2011ApJ...738..187L,2013A&A...554A..44A,2016A&A...596A..98A}, which suggests that thin disc stars with higher metallicities still have a guiding center radius smaller than that for the lower metallicity stars, and they have larger velocity dispersion and asymmetric drift. However, it is not clear that the velocity dispersion of intermediate-old thin disc stars decreases with radius \citep{2014ApJ...781L..20M}. 

This  paper  is  organized as  follows. In Section~\ref{data} we describe the stellar sample and how we break it up into chemical groups. Section~\ref{stellar} explains parallaxes, proper motions, radial velocities and how they are used to derive individual space velocities for the stellar sample. In Section~\ref{rot_vel} we explore the rotation-metallicity relation for the chemical groups. Section~\ref{ellipsoid} presents our analysis of the velocity ellipsoid, linear Str\"omberg relation and the Galactic velocity anisotropy parameter. The most important results are summarized and discussed  in  Section~\ref{summary}. 


\section{The data}
\label{data}

\subsection{The stellar sample}

We make use of the high-resolution spectroscopic study of 714 F and G intermediate-old dwarfs and subgiant stars in the solar neighbourhood by \citet{2014A&A...562A..71B}. The spectra have high resolution ($R$ = 40,000 - 110,000) and signal-to-noise (SNR = 150 - 300). They were obtained with the FEROS spectrograph on the ESO Very Large Telescope, the HARPS spectrograph on the ESO 3.6 m telescope, and the MIKE spectrograph on the Magellan Clay telescope. \citet{2014A&A...562A..71B} provided detailed elemental abundances for O, Na, Mg, Al, Si, Ca, Ti, Cr, Fe, Ni, Zn, Y and Ba. The determination of stellar parameters and elemental abundances is based on analysis using equivalent widths and one-dimensional, plane-parallel model atmospheres calculated under local thermodynamical equilibrium (LTE). Departures from the assumption of LTE (NLTE corrections) were applied to Fe I line, (see \citealt{2014A&A...562A..71B} for details.)

\subsection{The chemical groups}
\label{chem_grp}

\begin{figure}
  \centering  
  \includegraphics[width=1.05\columnwidth]{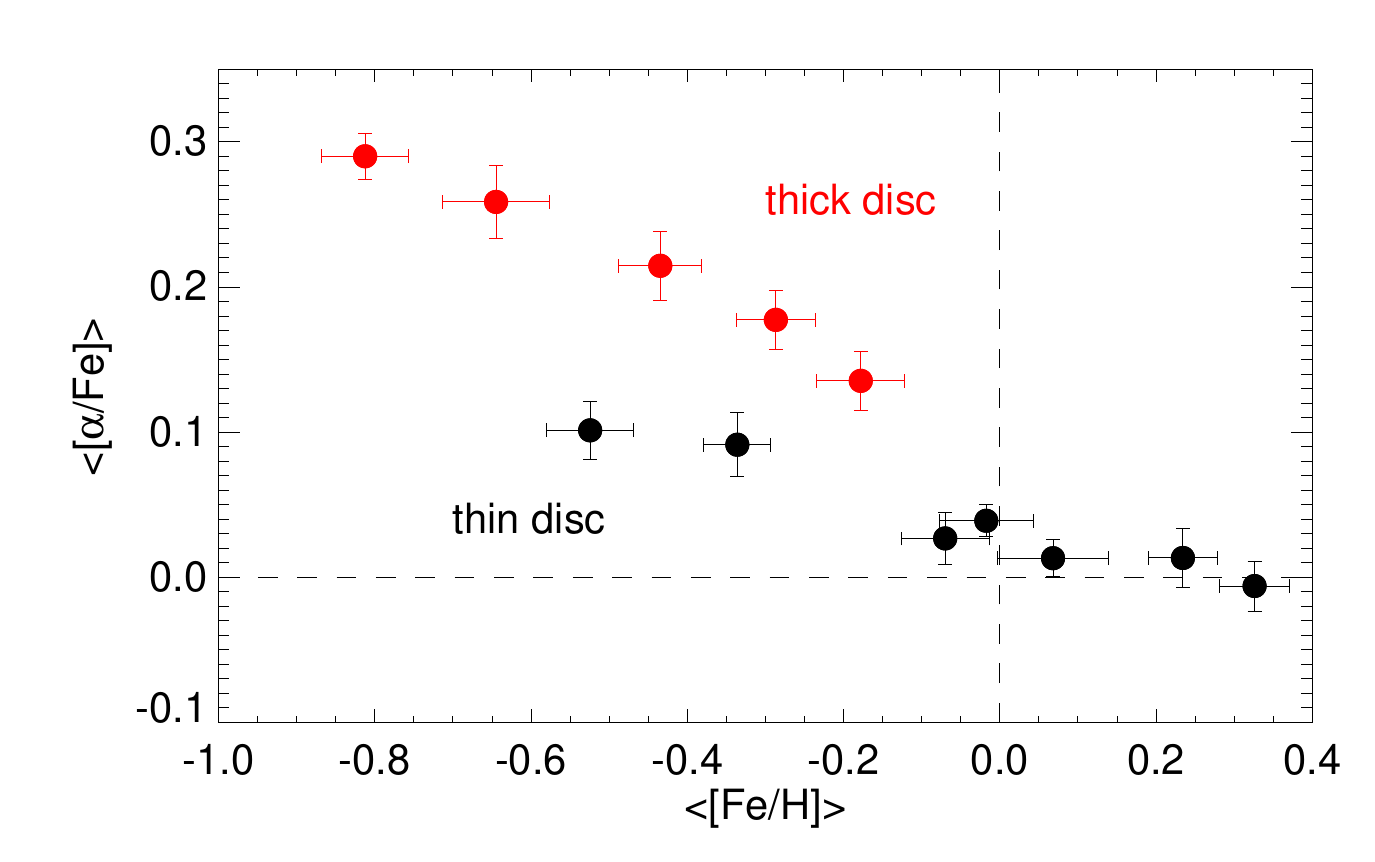}
   \caption{$<$[Fe/H]$>$ - $<$[$\alpha$/Fe]$>$ plane for the twelve chemical groups indentified in \citet{2014MNRAS.438.2753M} and employed in this study. The black dots are associated with the thin disc while the red dots to the thick Galactic disc. We respect this color-code through the entire paper.}
  \label{fig:al_fe_groups}
\end{figure}

The chemical tagging experiment by \citet{2013MNRAS.428.2321M, 2014MNRAS.438.2753M} used the sample described above to identify groupings of nearby disc field stars that share metal abundances using a Manhattan distance metric. The field stars they identified as having similar abundances are not clustered in space, nor do they share similar space motions \citep{2015MNRAS.450.2354Q}. These groups represented a first attempt to identify groups of stars from single, discrete birth events. However, it is beyond the scope of this paper to analyze whether the stars in a chemical group were born in the same molecular cloud or formed in different clusters but with the same chemical patterns to the level of precise abundance determination of \cite{2014A&A...562A..71B}. We refer the reader to \citet{2007AJ....133.1161D}, \citet{2014MNRAS.438.2753M}, \citet{2015A&A...577A..47B}, \citet{2016ApJ...833..262H}, \citet{2017arXiv170107829N} and references therein where this topic is debated in detail. In this study, we rather focus on the usefulness of chemical groups for exploring the kinematical properties of the very local thin and thick Galactic disc. 

\begin{table}
\caption{Number of stars in each chemical group (N) together with the average abundances values ($<$[Fe/H]$>$, $<$[$\alpha$/Fe]$>$) and standard deviations for the 12 chemical groups.}
\begin{center}
\begin{tabular}{cccc}
\hline
\hline
& N & $<$[Fe/H]$>$ (dex) & $<$[$\alpha$/Fe]$>$ (dex) \\
\hline
Thin disc & 41 & +0.07 $\pm$ 0.07 & +0.01 $\pm$ 0.01\\
               & 25 & --0.07 $\pm$ 0.06 & +0.03 $\pm$ 0.02 \\
	      & 21 & +0.23 $\pm$ 0.04 & +0.01 $\pm$ 0.02 \\
	      & 19 & --0.02 $\pm$ 0.06 & +0.04 $\pm$ 0.01 \\
	      & 16 & --0.52 $\pm$ 0.05 & +0.10 $\pm$ 0.02 \\
	      & 15 & --0.34 $\pm$ 0.04 & +0.09 $\pm$ 0.02 \\
	      & 15 &  +0.33 $\pm$ 0.04 & --0.01 $\pm$ 0.02 \\
	      \hline
Thick disc & 30 &  --0.29 $\pm$ 0.05 & +0.18 $\pm$ 0.02 \\
	         & 24 & --0.43 $\pm$ 0.05 & +0.21 $\pm$ 0.02 \\
	         & 21 & --0.64 $\pm$ 0.07 & +0.26 $\pm$ 0.02 \\
	         & 17 & --0.81 $\pm$ 0.05 & +0.29 $\pm$ 0.01 \\
	         & 15 & --0.18 $\pm$ 0.06 & +0.13 $\pm$ 0.02 \\
\hline
\end{tabular}
\end{center}
\label{tab:Fe_AL_ratio}
\end{table}%

From all the groups identified in \citet{2014MNRAS.438.2753M} we only select those where the number of stars is larger or equal to 15 (N $\geq$ 15), a total of twelve groups. Figure~\ref{fig:al_fe_groups} shows the ratio of the mean $\alpha$-elements to the iron abundance, $<$[$\alpha$/Fe]$>$, as a function of the mean iron abundance, $<$[Fe/H]$>$, for the selected chemical groups, and Table~\ref{tab:Fe_AL_ratio} shows their average abundance values and standard deviations together with the number of stars in each group. Different studies of abundance populations in the solar vicinity based on  high resolution spectroscopy find that the [$\alpha$/Fe] distribution is bimodal, suggesting that the thin and the thick disc are chemically distinguishable ---   see, e.g., \citet{2011MNRAS.412.1203N},  \citet{2011MNRAS.414.2893F}, \citet{2014ApJ...796...38N}, and \citet{2015MNRAS.453.1855M}. Using the twelve selected chemical groups we found at least two population sequences in the $<$[Fe/H]$>$ - $<$[$\alpha$/Fe]$>$ plane. One sequence, associated with the thin disc, has [$\alpha$/Fe] $\leq$ 0.1 dex and [Fe/H] ranging from +0.3 to --0.5 dex, while the second sequence, associated with the thick disc, has [$\alpha$/Fe] $>$ 0.1 dex and [Fe/H] ranging from --0.2 to --0.8 dex (see Fig.~\ref{fig:al_fe_groups}). In this work, we use the chemically distinguished thin and thick disc groups to explore the properties of the disc velocity ellipsoid and vertex deviation.
  
\section{Stellar kinematics}
\label{stellar}

\subsection{Parallaxes}

For the stellar parallaxes of the sample employed in this study we make use of the first \emph{Gaia} data release (\emph{Gaia} DR1), where the combination of positional information from the Hipparcos and \emph{Tycho}-2 catalogues allows the derivation of positions, parallaxes, and proper motions for about 2 million sources from the first 14 months of observations \citep{2016A&A...595A...2G}.  

\begin{figure}
  \centering  
  \includegraphics[width=1.04\columnwidth]{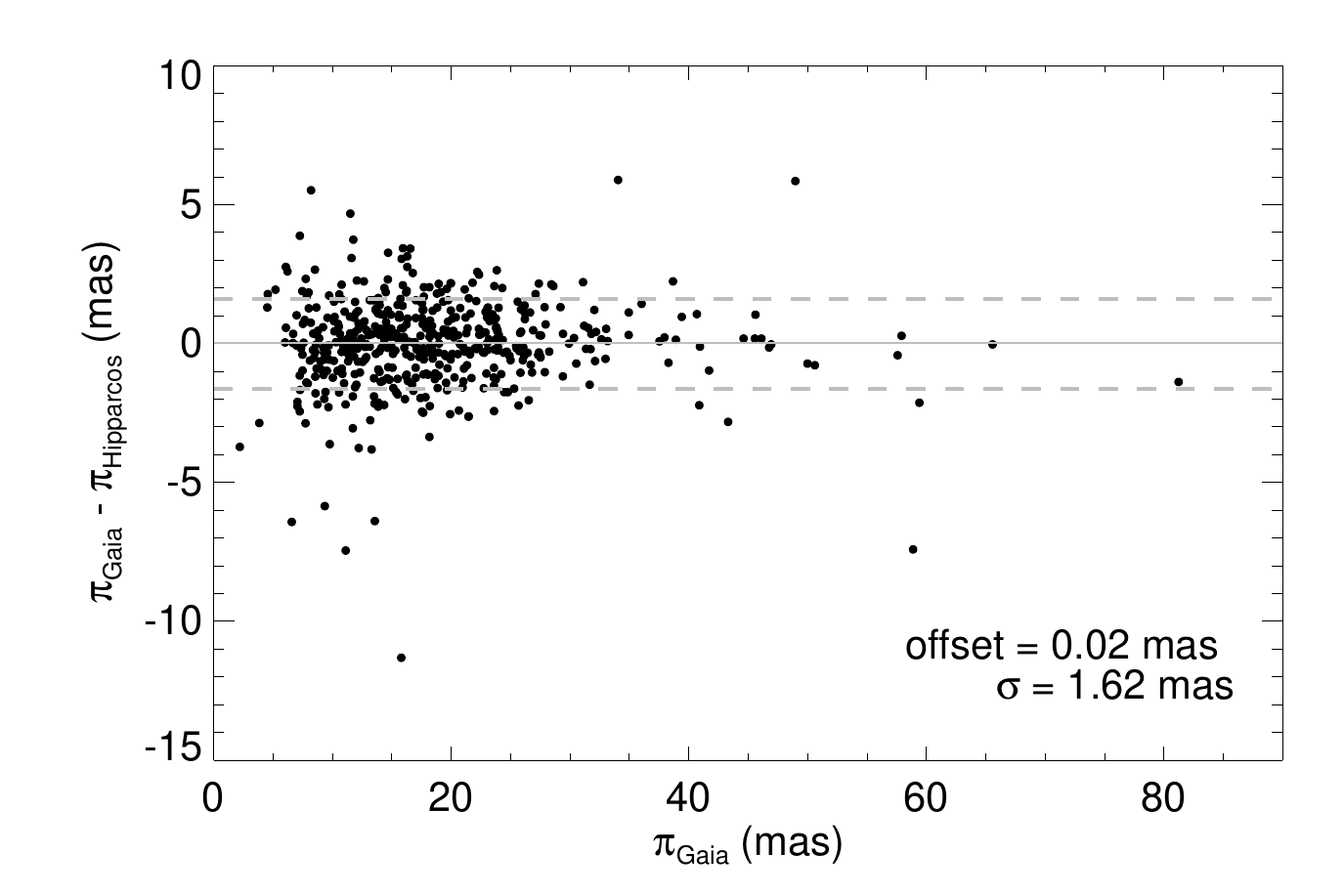}
     \caption{Parallaxes from \emph{Gaia} DR1 minus the parallaxes from Hipparcos as a function of the parallaxes from \emph{Gaia}. The grey line and the dashed grey line represents the mean difference and standard deviation of the entire sample, respectively.}
  \label{fig:plx_Gaia}
\end{figure}

In Figure~\ref{fig:plx_Gaia} we illustrate the difference in milliarcseconds (mas) between the parallaxes ($\pi$) measured from \emph{Gaia} DR1 and those measured in Hipparcos \citep{2007A&A...474..653V} as a function of the parallaxes from \emph{Gaia} for our stellar sample. There is no discernible offset between the measurements, and we find a small scatter of only 1.6 mas in the comparison. However, over large spatial scales it is known that the parallax zero-point variations reach an amplitude of $\sim$ 0.3 mas \citep{2016A&A...595A...4L}. \citet{2016A&A...595A...2G} stressed that a systematic component of $\sim$ 0.3 mas should be added to the parallax uncertainties. See also \cite{2017arXiv170901216B} for an update on the systematic errors in the Gaia DR1 parallaxes.

\begin{figure}
  \centering  
  \includegraphics[width=1.05\columnwidth]{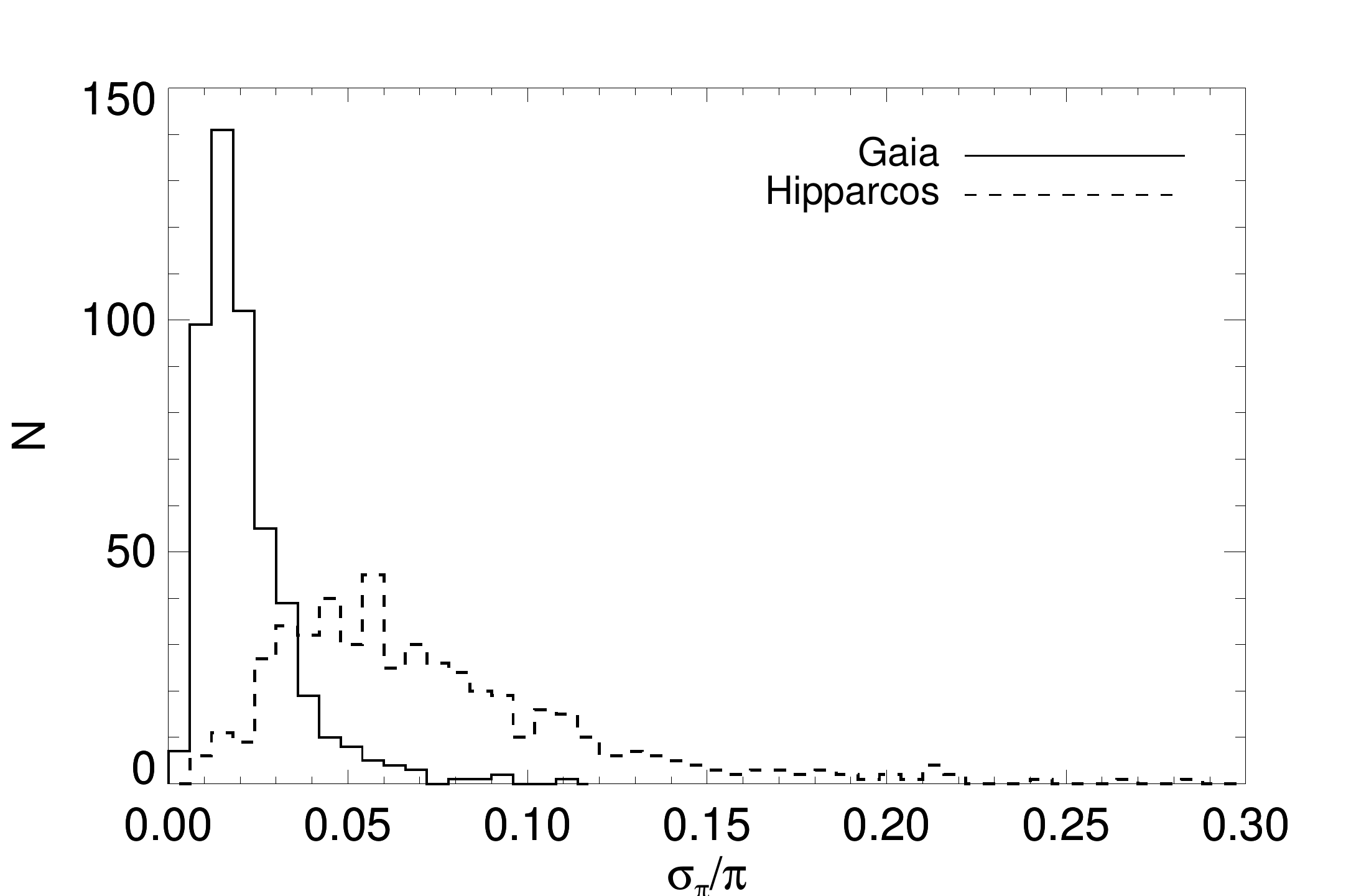}
   \caption{The distribution of the uncertainties in parallaxes from \emph{Gaia} DR1 (solid line) and Hipparcos (dashed line) for the stellar sample used in this study.}
  \label{fig:plx_Gaia_er}
\end{figure}

The distribution of the uncertainties in parallaxes from Hipparcos and \emph{Gaia} DR1 are shown in Figure~\ref{fig:plx_Gaia_er}. Nearly all the stars have a $(\sigma_{\pi}/\pi)_{Gaia}$ $<$ 0.07 while for most of the objects $(\sigma_{\pi}/\pi)_{Gaia}$ $\sim$ 0.01. For our stellar sample, $(\sigma_{\pi}/\pi)_{\rm Hipparcos}$ shows a broad distribution ranging from 0.01 to 0.30 (see dashed line in Fig.\,\ref{fig:plx_Gaia_er}).



\subsection{Proper motions and radial velocities}

\begin{figure*}
  \centering  
  \includegraphics[width=1.04\columnwidth]{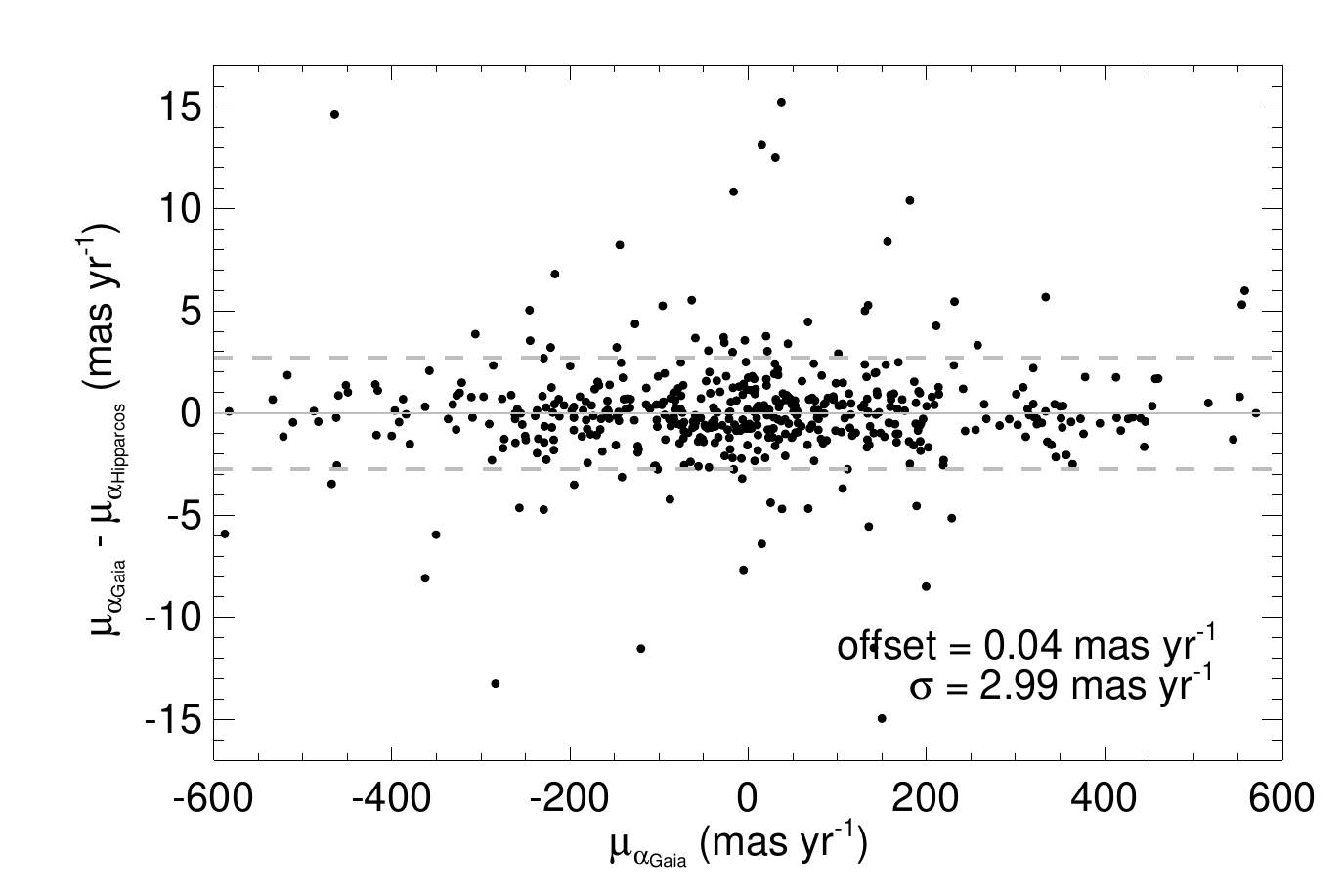}
  \includegraphics[width=1.04\columnwidth]{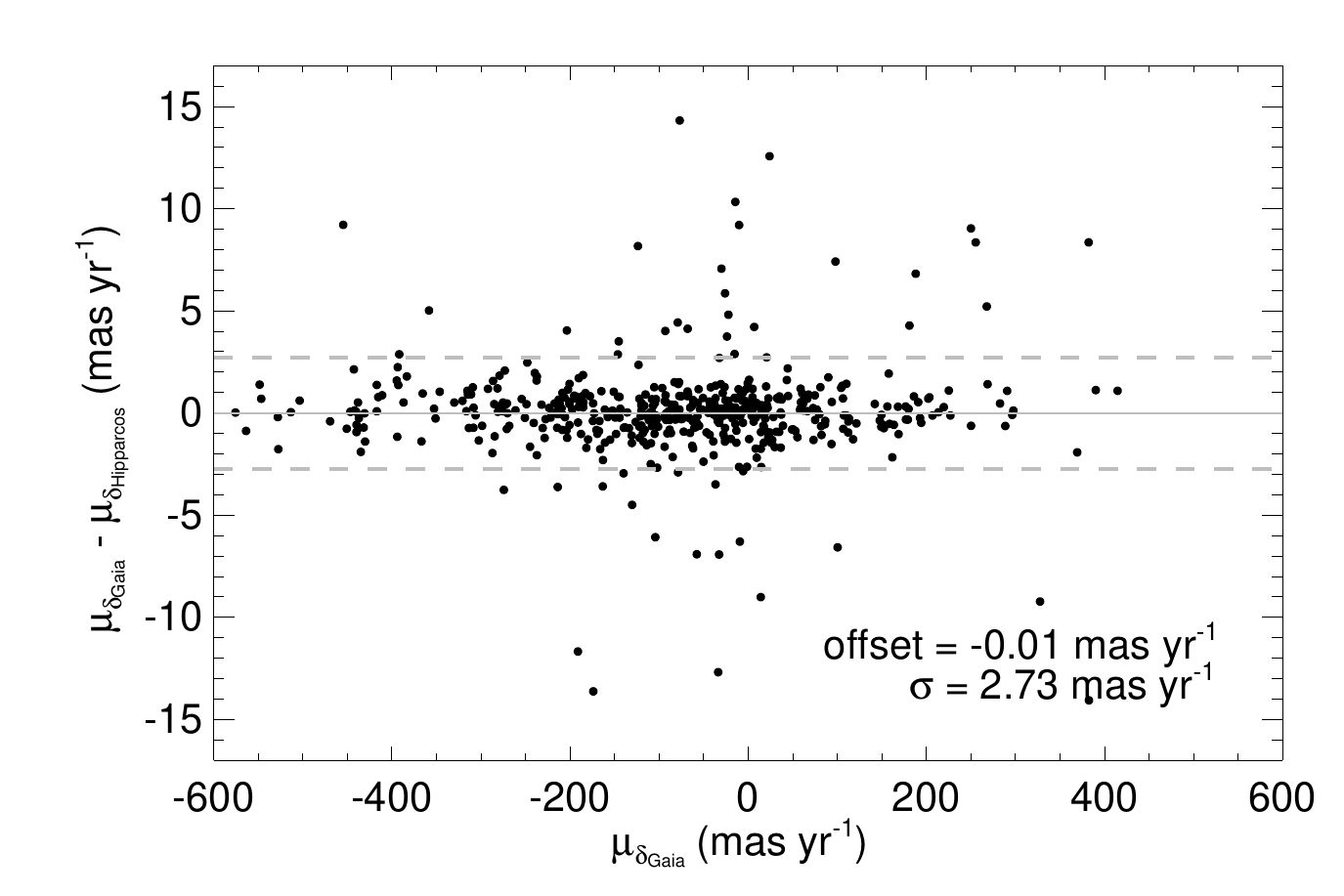} 
   \caption{Differences in the proper motions from \emph{Gaia} DR1 versus the proper motions from Hipparcos as a function of the proper motions from Gaia. In the left panel we have the right ascension proper motion component and in the right panel the declination proper motion component. The grey and dashed grey lines represent the mean difference and standard deviation, respectively, where the values are shown in the legend.}
  \label{fig:pmRADEC_Gaia}
\end{figure*}

\begin{figure}
  \centering  
  \includegraphics[width=1.05\columnwidth]{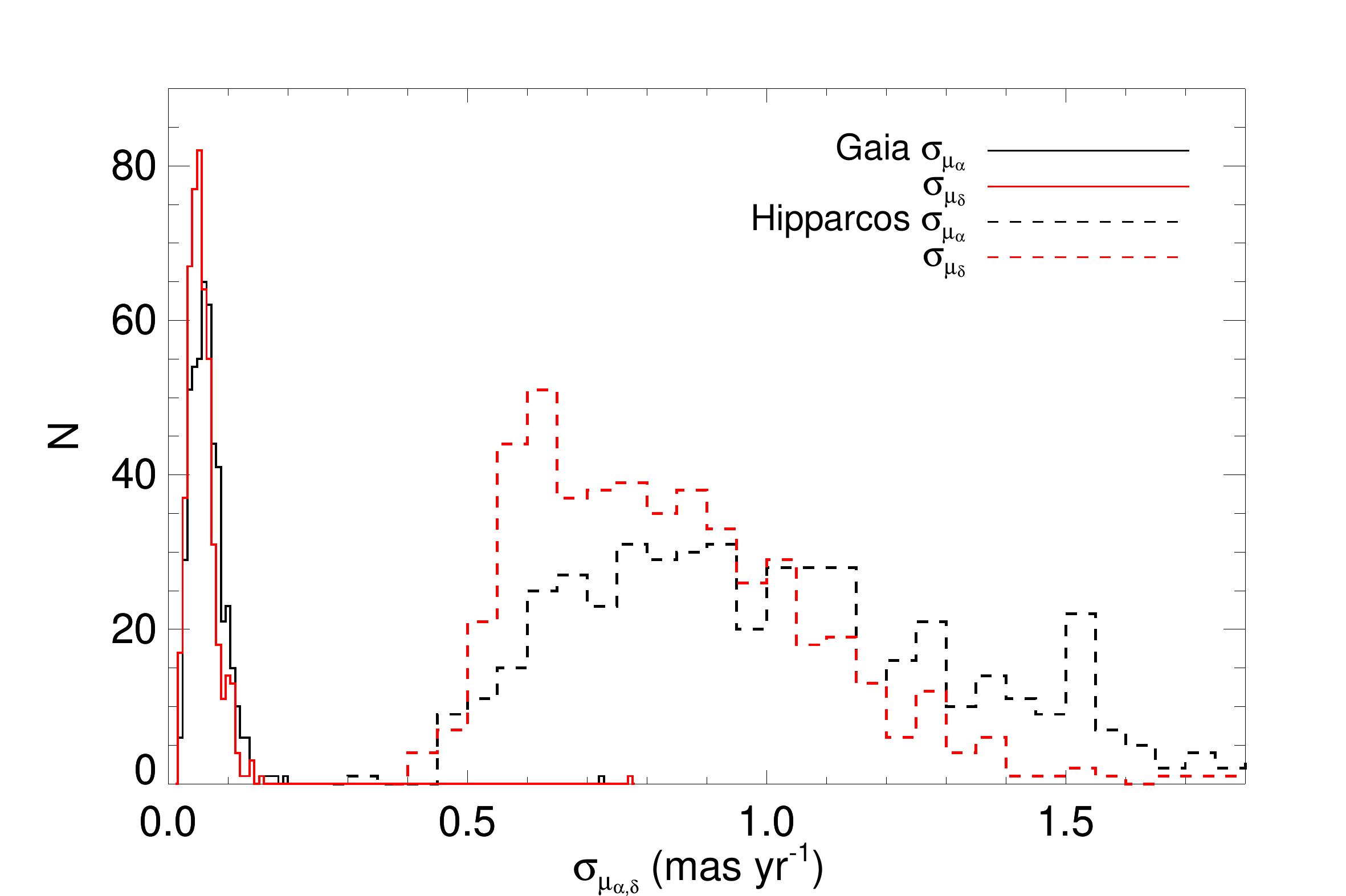} 
   \caption{Uncertainty distributions in proper motions from \emph{Gaia} DR1 (solid black and red lines) and from Hipparcos (dashed black and red lines). Nearly all the proper motions from \emph{Gaia} DR1 have a $\sigma_{\rm \mu_{\rm \alpha,\delta}}$ $<$ 0.2 mas yr$^{-1}$ with a peak around 0.05 mas yr$^{-1}$.}
  \label{fig:pm_error}
\end{figure}

We also make use of the high quality \emph{Gaia} DR1 proper motions realized by the Tycho-Gaia astrometric solution (TGAS), where the typical uncertainty is about 1 mas yr$^{-1}$ for the proper motions. The stellar sample we employ in this study is included in a subset of 93,635 bright Hipparcos stars in the primary astrometric data set where the proper motions are much more precise, at about 0.06 mas yr$^{-1}$ \citep{2016A&A...595A...2G}. In Figure~\ref{fig:pmRADEC_Gaia} we compare the measurement of the proper motion in right ascension and declination respectively between the Hipparcos and \emph{Gaia} DR1 catalogues. There is a scatter of 3 mas yr$^{-1}$ in $\mu_{\rm \alpha}$ and 2.7 mas yr$^{-1}$ in $\mu_{\rm \delta}$ when comparing both measurements. Figure~\ref{fig:pm_error} clearly shows the major improvement in astrometry from \emph{Gaia} DR1 with respect to Hipparcos. All the stars in this sample have a $\sigma_{\rm \mu_{\rm \alpha,\delta}}$ $<$ 0.2 mas yr$^{-1}$ from \emph{Gaia} DR1 (solid black and red lines) while the uncertainties in proper motions from Hipparcos show values ranging from 0.4 to 2.0 mas yr$^{-1}$ (dashed black and red lines in Fig.\,\ref{fig:pm_error}). Most of our stars show a $\sigma_{\rm \mu_{\rm \alpha,\delta}}$ $\sim$ 0.05 mas yr$^{-1}$. A systematic uncertainty at about 0.06 mas yr$^{-1}$ may be present in the proper motions from \emph{Gaia} DR1 \citep{2016A&A...595A...2G}.   

The radial velocities (RVs) employed in this study were obtained from the Geneva-Copenhagen Survey (GCS), using the photoelectric cross-correlation spectrometers CORAVEL (see \citealt{2004A&A...418..989N} for details). All stars have uncertainties in RVs better than 0.7 km s$^{-1}$.  

\subsection{Space velocities}

\begin{figure}
  \centering  
  \includegraphics[width=1.05\columnwidth]{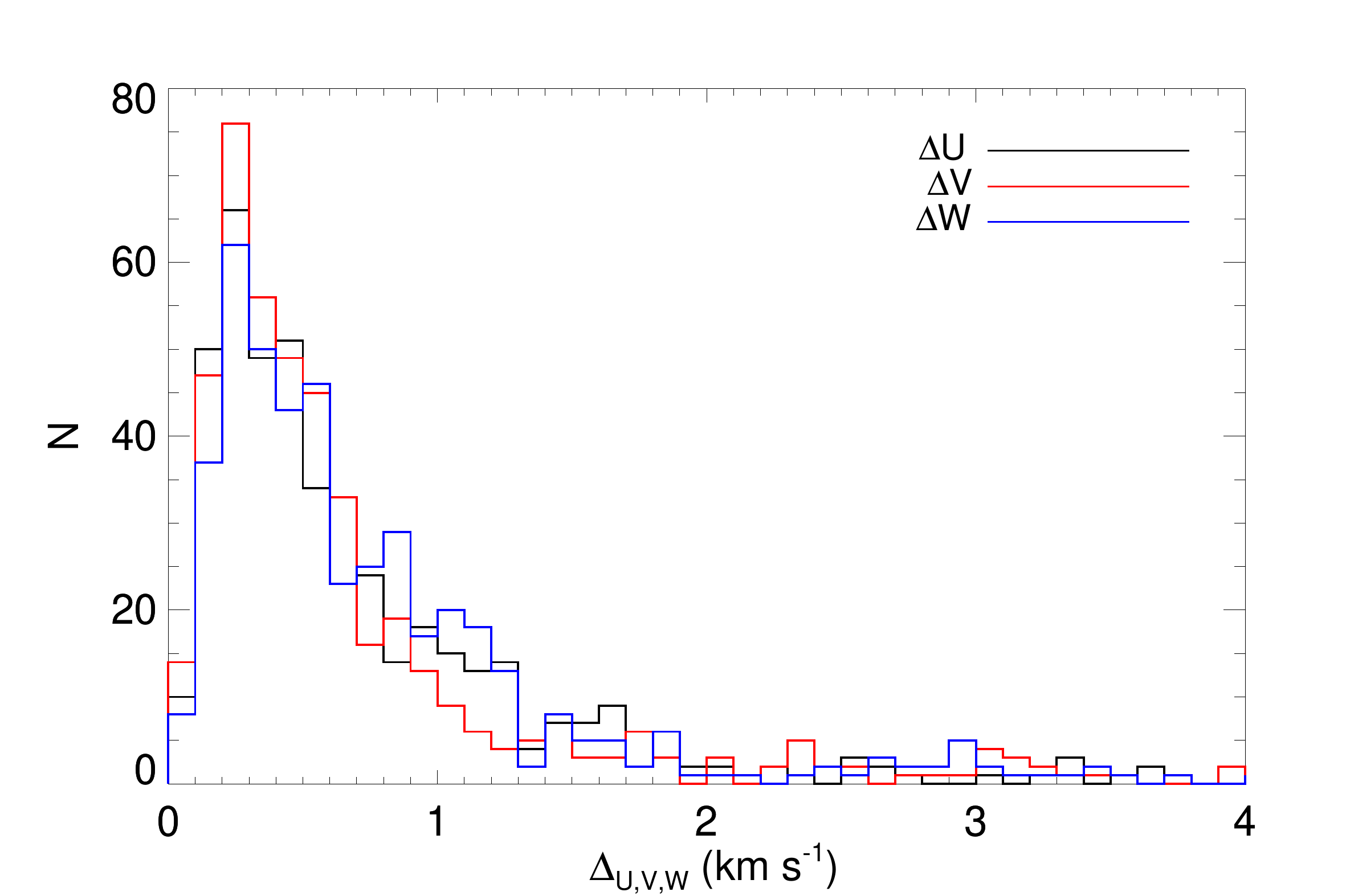} 
   \caption{Random uncertainties distributions for the three components of the space velocities (U,V,W) using parallaxes and proper motions from \emph{Gaia} DR1. The typical uncertainty in velocity is 0.4 km s$^{-1}$ and most of the stellar sample has $\Delta$U, $\Delta$V, $\Delta$W $<$ 1.3 km s$^{-1}$.}
  \label{fig:velo_error}
\end{figure}

Velocities in a Cartesian Galactic system were obtained following the equations in \citet{1987AJ.....93..864J}. That is, from the observed GCS radial velocities, and the \emph{Gaia} DR1 proper motions and parallaxes, we derive the space velocity components (U, V, W). We adopt a right-handed Galactic system, where U is pointing towards the Galactic centre, V in the direction of rotation, and W towards the North Galactic Pole (NGP). The uncertainties in the velocity components U , V and W were derived also using the formalism introduced by \citet{1987AJ.....93..864J}. The sources of uncertainties are the distances, the proper motions and the radial velocities where the errors of these measured quantities are uncorrelated, i.e., the covariances are zero. We illustrate in Figure~\ref{fig:velo_error} the random uncertainties distribution for the three stellar velocity components, i.e., $\Delta$U, $\Delta$V, $\Delta$W. The typical uncertainties for the space velocities in the stellar sample is just 0.4 km s$^{-1}$.  



\begin{figure*}
  \centering  
  \includegraphics[width=2.15\columnwidth]{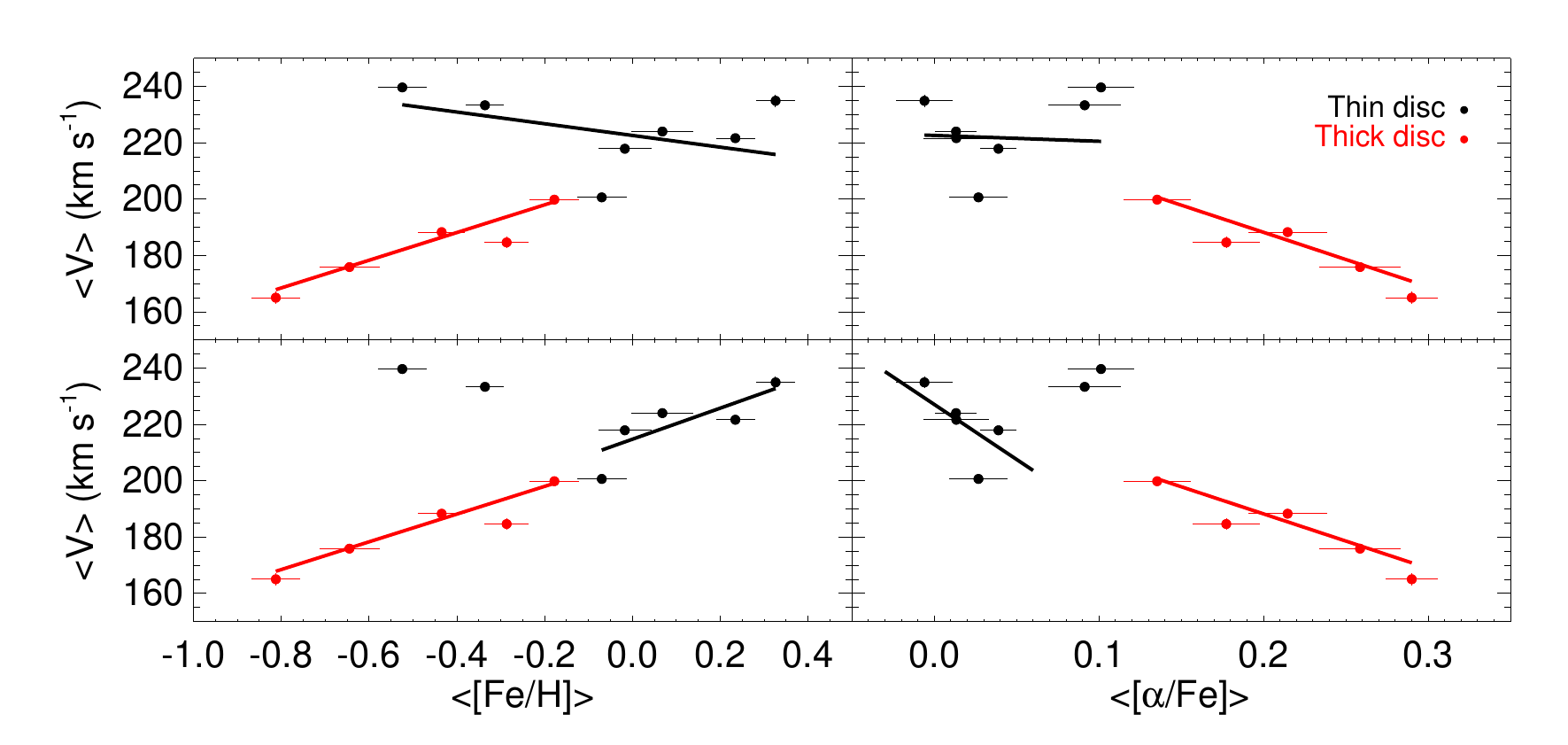}
     \caption{The Galactic rotational velocity derived using Gaia DR1 parallaxes and proper motions for the twelve chemical groups in the thin disc (black dots) and thick disc (red dots) as a function of $<$[Fe/H]$>$ and $<$[$\alpha$/Fe]$>$, respectively. The top-panels show the best linear fit using all the thin disc chemical groups (black line) and thick disc groups (red line). Alternatively, in the bottom-panels we show the best fit excluding the most metal-poor thin disc groups (black line), where a different trend appears. See text for details and Sec.~\ref{summary} for discussion.}
  \label{fig:V_Fe}
\end{figure*}



\begin{figure*}
  \centering  
  \includegraphics[width=2.1\columnwidth]{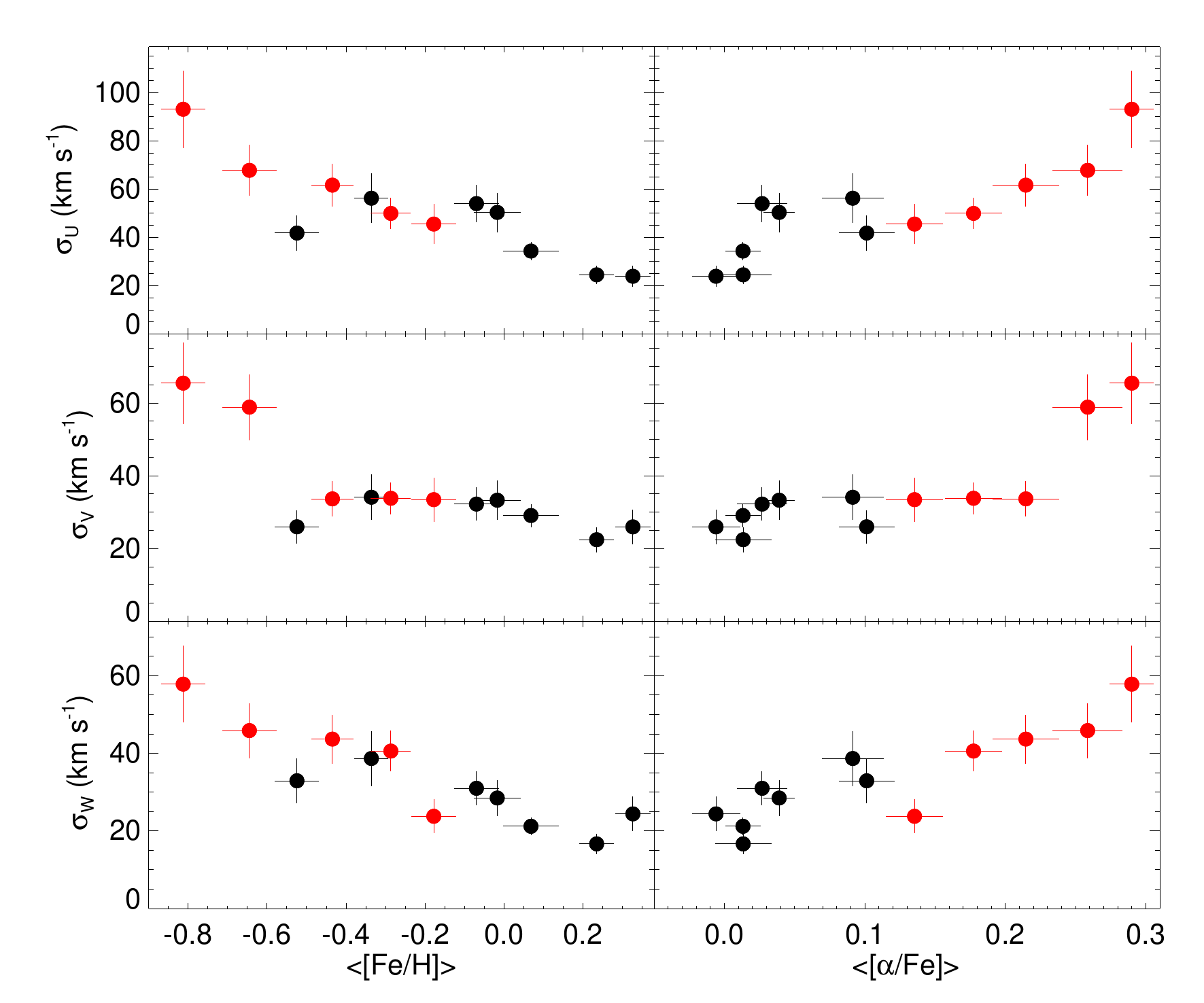}
   \caption{Velocity dispersion for the three velocity components as a function of the mean iron abundances and $\alpha$-elements for the twelve chemical groups. The black dots are associated with the thin disc, the red to the thick disc. In all cases it appears that the kinematics do not clearly discriminates between thin and thick disc in their trends. The trends smoothly overlap and extend each other.}
  \label{fig:al_fe_disper}
\end{figure*}

\begin{figure*}
  \centering  
  \includegraphics[width=2.1\columnwidth]{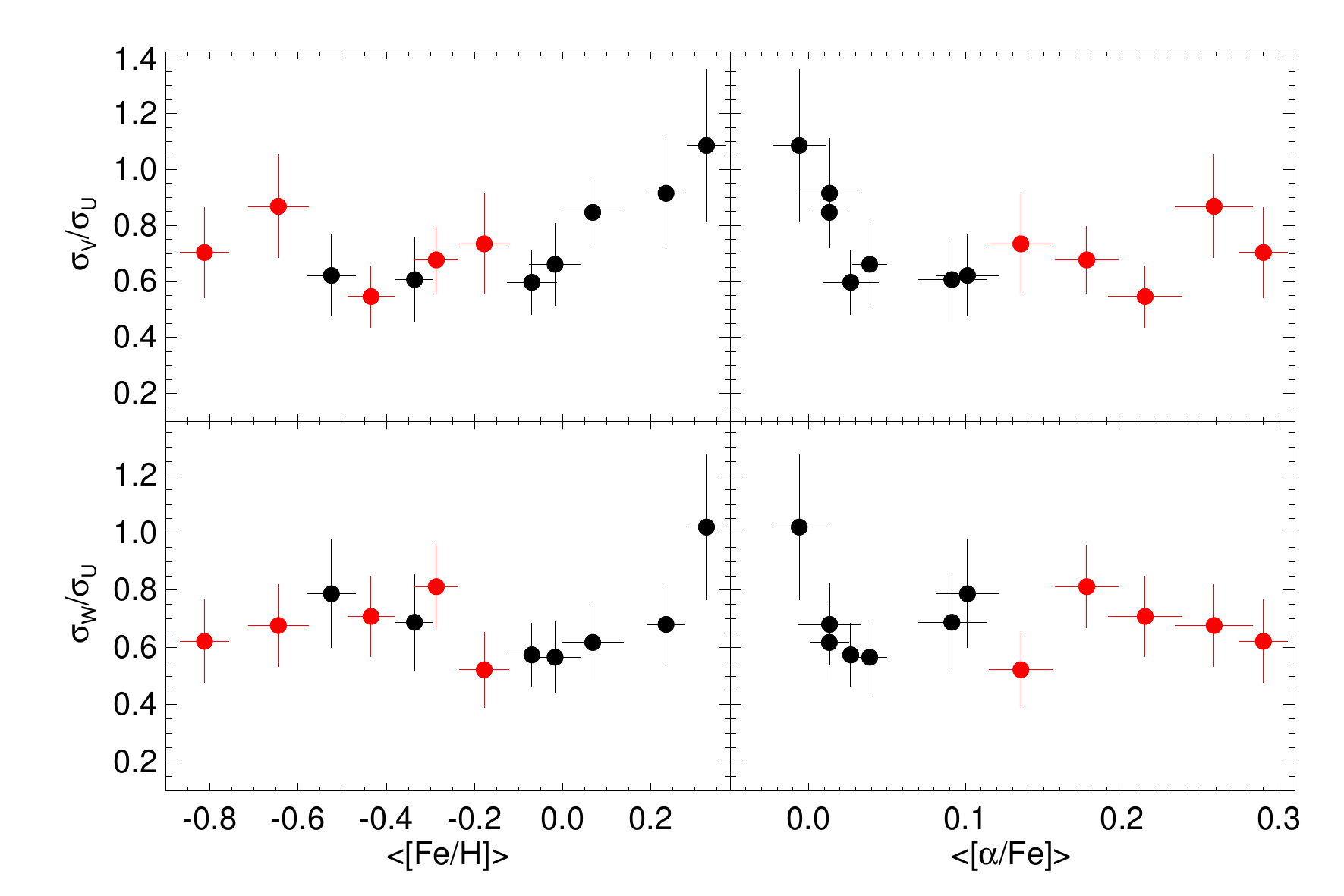}
   \caption{Ratio of dispersions as a function of $<$[Fe/H]$>$ and $<$[$\alpha$/Fe]$>$ for the chemical groups analyzed here. The upper panel represents $\sigma_{\rm V}/\sigma_{\rm U}$, the lower panel $\sigma_{\rm W}/\sigma_{\rm U}$.}
  \label{fig:al_fe_ellip}
\end{figure*}

\section{The rotation-abundance relation}
\label{rot_vel}

In this section we use the chemical groups to examine the correlations between the rotational velocity and the mean abundances of the groups. Figure~\ref{fig:V_Fe} shows the relationship between the mean Galactic rotation\footnote{The LSR was assumed to be on a circular orbit, with circular velocity
$\Theta = 240 \pm  8$~km~s$^{-1}$ \citep{2014ApJ...783..130R}.} and $<$[Fe/H]$>$ and $<$[$\alpha$/Fe]$>$, respectively. We find a clear correlation between the $<$V$>$ and the abundances for the thick disc groups (red points): the asymmetric drift increases as the groups get more metal-poor and more $\alpha$-enhance. However, for the thin disc any correlation is less clear. The top-left panel in Figure~\ref{fig:V_Fe} shows the best linear fit (black line) in the $<$V$>$-[Fe/H] diagram using all seven thin disc groups. The trend shows an opposite sign for the thin and the thick disc, where more metal-poor thin disc stars have the smallest (near zero) asymmetric drift. This result is consistent with other studies addressing this problem, e.g., \citet{2011ApJ...738..187L}, and \citet{2016A&A...596A..98A}. Interestingly, there is a group with thin disc abundances (see Fig.~\ref{fig:al_fe_groups} and Fig.~\ref{fig:V_Fe}) but that lies in the thick disc trend in $<$V$>$ - [Fe/H]. However, the behaviour described above is less clear in the $<$V$>$-[$\alpha$/Fe] diagram for the thin disc groups, where there is no evident correlation between the velocity and the $\alpha$-elements. 

We adopt an alternative view for the bottom panels of Figure~\ref{fig:V_Fe}. The most metal-poor chemical groups in the thin disc show a different trend with respect most of the groups (Fig.~\ref{fig:V_Fe}). When excluding them from the linear fit, we find a tight relation between the $<$V$>$ and [Fe/H], where metal-poor groups lag behind with respect to the metal-rich ones, in the same manner as for the thick disc groups. A similar result is found in the $<$V$>$-[$\alpha$/Fe] diagram (bottom-right in Fig.~\ref{fig:V_Fe}). We discuss the implications of these results in Section~\ref{summary}. 

\section{The velocity ellipsoid}
\label{ellipsoid}

The kinematics of a stellar population can be described by its velocity dispersion tensor

\begin{equation}
\sigma^{2}_{ij} = \langle(\upsilon_{i} - \langle\upsilon_{i}\rangle)(\upsilon_{j} - \langle\upsilon_{j}\rangle)
\end{equation}

where the subscript indices denote one of the orthogonal coordinate directions and the angled brackets represent averaging over the phase-space distribution function \citep{2008gady.book.....B, 2009ApJ...698.1110S}. We represent the velocity dispersion ($\sigma_{\rm U}$,$\sigma_{\rm V}$,$\sigma_{\rm W}$) of the twelve chemical groups as a function of $<$[Fe/H]$>$ and $<$[$\alpha$/Fe]$>$ in Figure~\ref{fig:al_fe_disper}. The error bars of the  velocity dispersion were computed as 
\begin{equation}
\Delta{V_{\rm i}} = (2N)^{-1/2} \sigma_{\rm i}
\end{equation}
where $N$ is the number of stars in each chemical group (see Tab.~\ref{tab:Fe_AL}). The error bars of the abundances are adopted as the standard deviations for each chemical group. In Fig.~\ref{fig:al_fe_disper} the black dots represent chemically selected thin disc groups while red dots are thick disc groups (Sect.~\ref{chem_grp}). Most of the thick Galactic disc population is kinematically hotter than that of the thin disc \citep{2012ASSP...26..137F, 2016ARA&A..54..529B}. We find that very metal-poor and $\alpha$-enhancement groups are typically hotter than their more metal-rich and low $\alpha$-enhancement counterparts though, not all of the chemically selected thick disc chemical groups are hotter than the thin disc ones. 

\begin{table}
\caption{Average velocity dispersion values for the three space velocity components for the thin and thick disc.}
\begin{center}
\begin{tabular}{cccc}
\hline
\hline
& $<\sigma_{U}>$ & $<\sigma_{V}>$ & $<\sigma_{W}>$ \\
\hline
Thin disc & 33 $\pm$ 4 km s$^{-1}$ & 28 $\pm$ 2 km s$^{-1}$ & 23 $\pm$ 2 km s$^{-1}$ \\
Thick disc & 57 $\pm$ 6 km s$^{-1}$ & 38 $\pm$ 5 km s$^{-1}$ & 37 $\pm$ 4 km s$^{-1}$ \\
\hline
\end{tabular}
\end{center}
\label{tab:Fe_AL}
\end{table}%

For the radial velocity component we find that $\sigma_{\rm U}$ ranges from 25 to $\sim$ 50 km s$^{-1}$ for the thin disc chemical groups. The radial velocity dispersion increases from $<$[Fe/H]$>$ = +0.3 dex to $<$[Fe/H]$>$ = -0.1 dex and then remains nearly constant from -0.1 to -0.5 dex (black dots in Fig.~\ref{fig:al_fe_disper} top-left panel). Using a weighted mean where the weights are given by the combined error associated with the three components of the velocity, we obtain an average value for the thin disc groups of $\sigma_{\rm U}$ = 33 $\pm$ 4 km s$^{-1}$. In the case of the thick disc chemical groups (red dots in Fig.~\ref{fig:al_fe_disper} top-left panel), $\sigma_{\rm U}$ ranges from 40 to 70 km s$^{-1}$ and there is one group with $<$[Fe/H]$>$ $\sim$ -0.85 dex and $\sigma_{\rm U}$ $\sim$ 95 km s$^{-1}$, suggesting that some of these stars could be part of the Galactic halo. The average value for the thick disc groups is $\sigma_{\rm U}$ = 57 $\pm$ 6 km s$^{-1}$. We also plot $\sigma_{\rm U}$ as a function of $<$[$\alpha$/Fe]$>$ (Fig.~\ref{fig:al_fe_disper} top-right panel). The selected thin disc groups range from $<$[$\alpha$/Fe]$>$ = -0.05 to +0.1 dex (black dots) while the thick disc groups range from +0.1 to +0.3 dex (red dots), following the selection criteria discussed in Section~\ref{chem_grp}. For the thin disc $\sigma_{\rm U}$ clearly increases from -0.05 to +0.05 dex and then rises within the errors. The radial velocity dispersion for the thick disc chemical groups increases constantly from +0.1 to +0.3 dex. 

The middle panels in Figure~\ref{fig:al_fe_disper} represent $\sigma_{\rm V}$ as a function of $<$[Fe/H]$>$ and $<$[$\alpha$/Fe]$>$. Interestingly, for the thin disc groups $\sigma_{\rm V}$ slightly increases from $\sim$ 20 to 30 km s$^{-1}$ in the +0.3 dex to 0.0 dex metallicity range and 0.0 to +0.05 dex in $<$[$\alpha$/Fe]$>$, and then remains nearly constant within the errors for the thin disc groups. We obtain an average value for the thin disc of $\sigma_{\rm V}$ = 28 $\pm$ 2 km s$^{-1}$. For the thick disc groups we also find that $\sigma_{\rm V}$ remains constant from -0.15 to -0.6 dex and from +0.1 to +0.2 dex for $<$[$\alpha$/Fe]$>$, where $\sigma_{\rm V}$ $\sim$ 35 km s$^{-1}$. For the two most metal-poor groups there is an abrupt increase of the velocity dispersion in the V component (red dots in Fig.~\ref{fig:al_fe_disper} middle panel), reaching values of $\sigma_{\rm V}$ $\sim$ 60 km s$^{-1}$. The average value for the thick disc groups is $\sigma_{\rm V}$ = 38 $\pm$ 5 km s$^{-1}$. 

The bottom panels in Fig.~\ref{fig:al_fe_disper} represent $\sigma_{\rm W}$ as a function of $<$[Fe/H]$>$ and $<$[$\alpha$/Fe]$>$. The value of $\sigma_{\rm W}$ increases with decreasing metallicity from 15 to almost 30 km s$^{-1}$ for the metal-rich groups in the thin disc while for the two metal-poor thin disc groups $\sigma_{\rm W}$ remains around 30 - 40 km s$^{-1}$. We obtain an average value of $\sigma_{\rm W}$ = 23 $\pm$ 2 km s$^{-1}$ for the thin disc groups, ranging from 0.0 to 0.1 dex in $<$[$\alpha$/Fe]$>$. The vertical velocity dispersion for the thick disc groups clearly increases when $<$[Fe/H]$>$ decreases and $<$[$\alpha$/Fe]$>$ increases, ranging from $\sim$ 20 to 45 km s$^{-1}$ while the most metal-poor group and $\alpha$-enhancement shows a $\sigma_{\rm W}$ $\sim$ 55 km s$^{-1}$. The average value for all thick disc groups is $\sigma_{\rm W}$ = 37 $\pm$ 4 km s$^{-1}$. 

\begin{table}
\caption{Weighted average and standard deviation velocity ellipsoid values for the thin and thick disc.}
\begin{center}
\begin{tabular}{cccc}
\hline
\hline
&  $<\sigma_{V}/\sigma_{U}>$ & $<\sigma_{W}/\sigma_{U}>$ \\
\hline
Thin disc  &  0.70 $\pm$ 0.13 & 0.64 $\pm$ 0.08 \\
Thick disc & 0.67 $\pm$ 0.11  & 0.66 $\pm$ 0.11 \\
\hline
\end{tabular}
\end{center}
\label{tab:ellip_weight}
\end{table}%


In Figure~\ref{fig:al_fe_ellip}, we represent the value of the ratio between the semi-axes of the velocity ellipsoid, $\sigma_{\rm V}$/$\sigma_{\rm U}$ and $\sigma_{\rm W}$/$\sigma_{\rm U}$ versus $<$[Fe/H]$>$ and $<$[$\alpha$/Fe]$>$ for the selected thin and thick disc chemical groups. These ratios are related to secular heating processes (e.g., \citealt{2016MNRAS.462.1697A} and references therein). For the thin disc groups (black dots), there is a trend between $\sigma_{\rm V}$/$\sigma_{\rm U}$ and $<$[Fe/H]$>$ and $<$[$\alpha$/Fe]$>$ ranging from 1.1 to 0.6, except for the groups with $<$[Fe/H]$>$ $<$ --0.2 and $<$[$\alpha$/Fe]$>$ $\sim$ 0.05 dex, where $\sigma_{\rm V}$/$\sigma_{\rm U}$ $\sim$ 0.6. For the very metal-rich groups $\sigma_{\rm V}$/$\sigma_{\rm U}$ is larger than for the rest of the thin disc chemical groups, however also the error bars are larger compared with the rest of the thin disc groups. The weighted average value for the thin disc is 0.70 $\pm$ 0.13 while similar value is found for the thick disc, 0.65 $\pm$ 0.13. We do not find any clear trend for the thick chemical groups between $\sigma_{\rm V}$/$\sigma_{\rm U}$ and the mean abundances (red dots). We also find that $\sigma_{\rm W}$/$\sigma_{\rm U}$ is nearly constant for the thin and thick disc groups with respect to the mean [Fe/H] and [$\alpha$/Fe] abundances (see bottom panels in Fig.~\ref{fig:al_fe_ellip}). For the most metal-rich group where $<$[Fe/H]$>$ and $<$[$\alpha$/Fe]$>$ are +0.33 and -0.01 dex, respectively, we find that $\sigma_{\rm W}$/$\sigma_{\rm U}$ = 1.0 $\pm$ 0.2, while for the rest of the groups $\sigma_{\rm W}$/$\sigma_{\rm U}$ is around 0.6. The average value for the thin disc chemical groups is 0.67 $\pm$ 0.11 and for the thick disc groups $<$$\sigma_{\rm W}$/$\sigma_{\rm U}$$>$ = 0.66 $\pm$ 0.11, again a similar value for both discs. We summary these results in Table~\ref{tab:ellip_weight}.

\begin{figure}
  \centering  
  \includegraphics[width=1.\columnwidth]{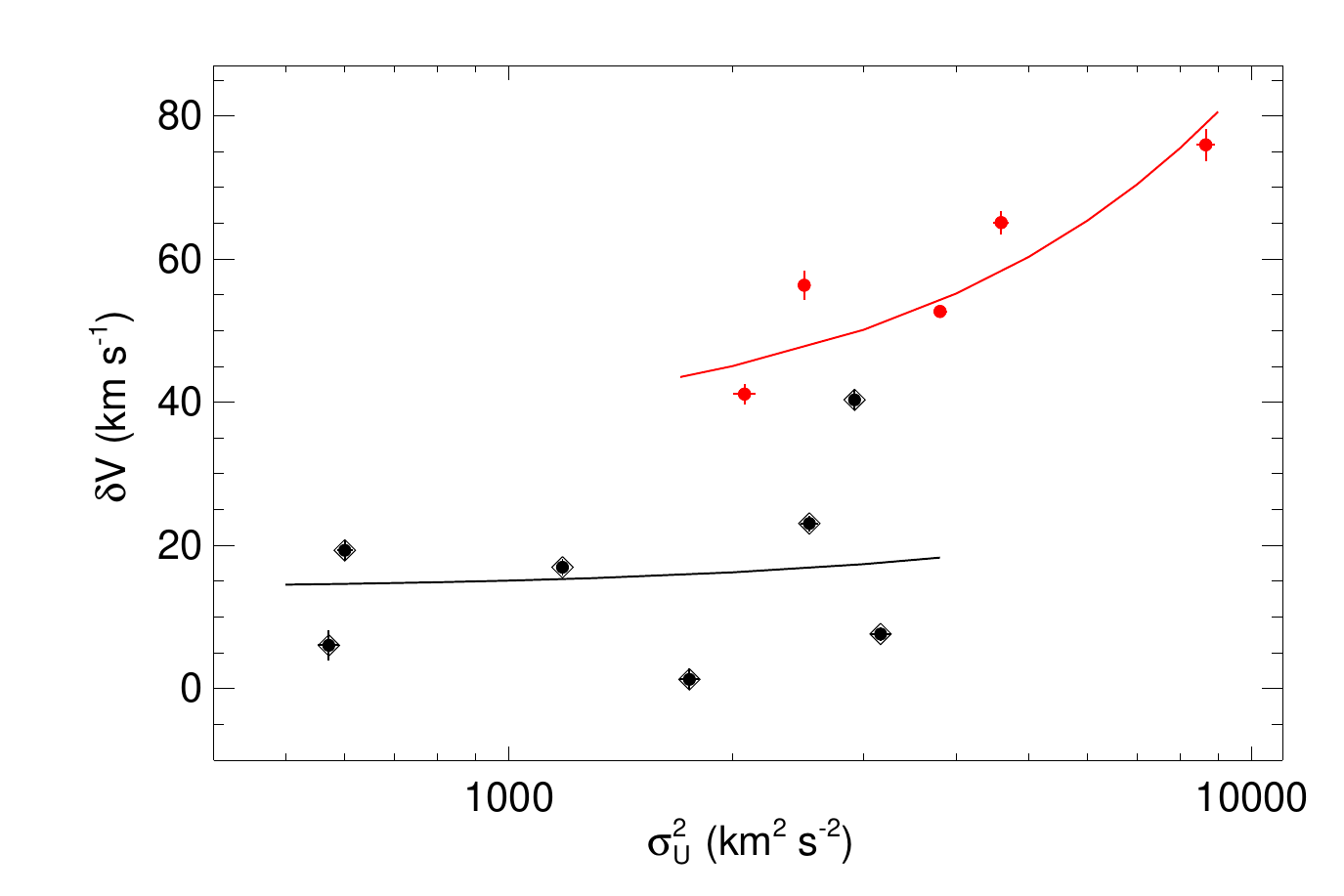}
   \caption{Asymmetric drift for the thin (black dots) and thick (red dots) chemical groups as a function of $\sigma_{\rm U}^{2}$. The black line gives the best linear fit to the thin disc data points, it corresponds to the component of the Sun's velocity with respect to the LSR, V$_{\sun}$ = 13.9 $\pm$ 3.4 km s$^{-1}$. The red line is the best linear fit for the thick disc groups.}
  \label{fig:stromberg_relation}
\end{figure}

\begin{figure*}
  \centering  
  \includegraphics[width=2.1\columnwidth]{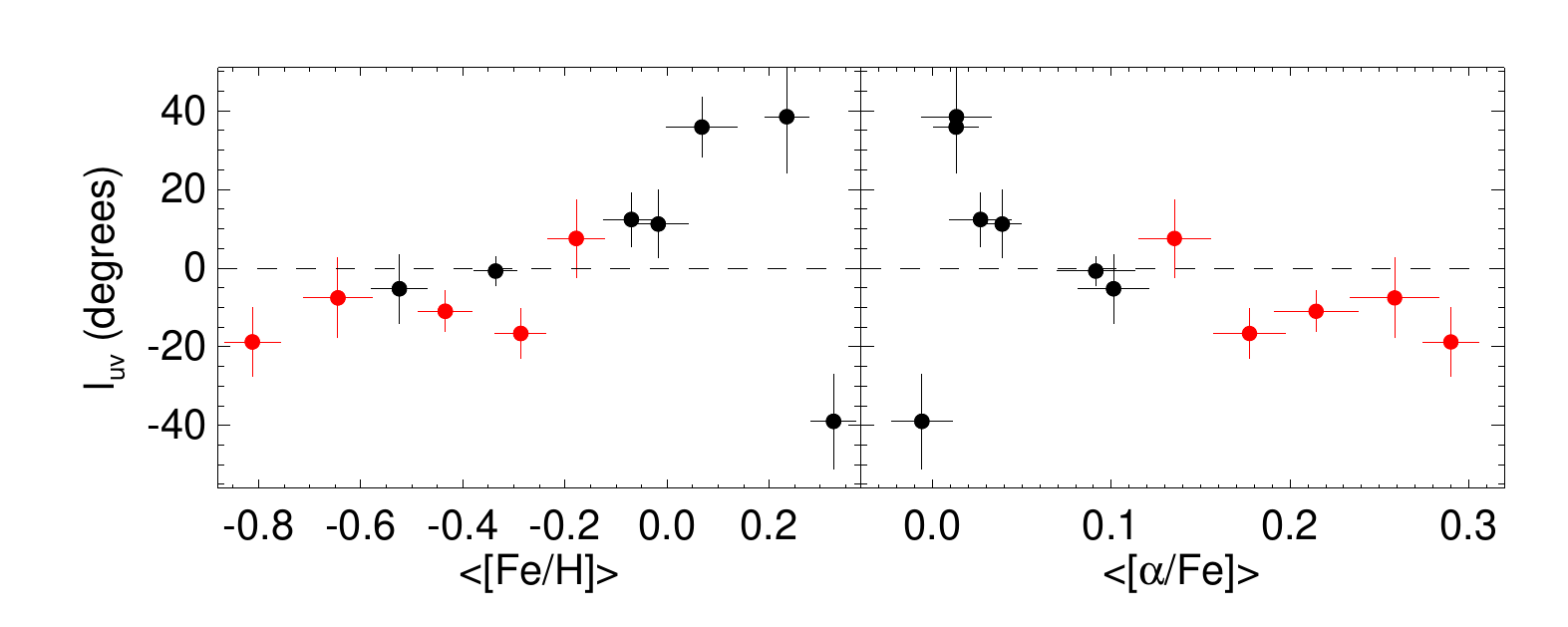}
   \caption{Measured vertex deviation as a function of abundances, [Fe/H] and [$\alpha$/Fe], respectively. We find a relation between l$_{\rm uv}$ and $<$[Fe/H]$>$ for the thin disc groups (black dots), while the vertex deviation for most of the thick disc groups is independent of the abundances (red dots).}
  \label{fig:fe_al_vertex}
\end{figure*}

\begin{figure*}
  \centering  
  \includegraphics[width=2.1\columnwidth]{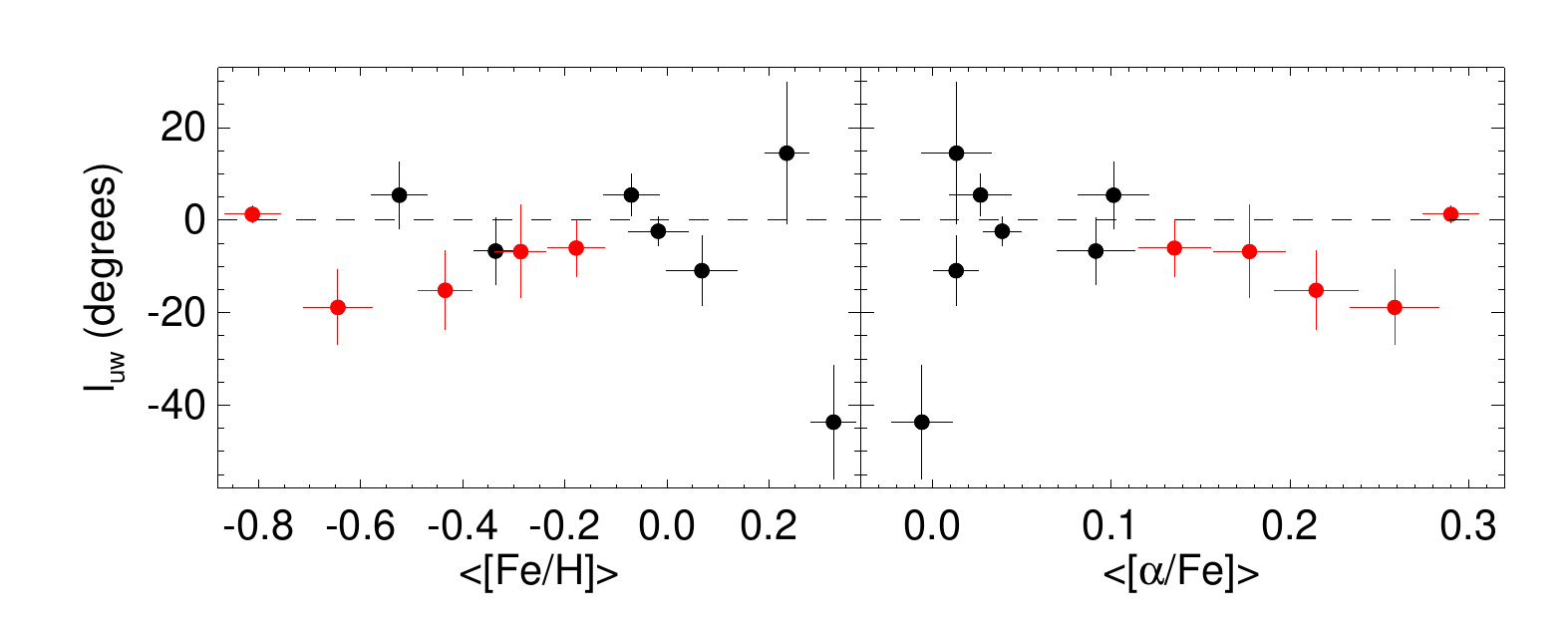}
   \caption{Tilt of the velocity ellipsoid as a function of iron abundances and $\alpha$-element abundances. There is no evident relation between the tilt angle and the abundances for the thin disc. A small non zero trend between the abundances of the thick disc groups and l$_{\rm uw}$ appears for most of them, excluding the most metal-poor group.}
  \label{fig:fe_al_tilt}
\end{figure*}

\subsection{The linear Str\"omberg relation}

\begin{table*}
\caption{Space velocity V$_{\sun}$ with respect to the LSR as given in the literature using the the Str\"omberg relation}
\begin{center}
\begin{tabular}{ccc}
\hline
\hline
Reference &  Source & V$_{\sun}$ (km s$^{-1}$)\\
\hline
This work  & \citet{2014MNRAS.438.2753M} & 13.9 $\pm$ 3.4 \\
Sperauskas et al. (2016) & McCormick K-M dwarfs & 14.2 $\pm$ 0.8 \\
 \citet{2010MNRAS.403.1829S} & Hipparcos &  12.2 $\pm$ 0.5 \\
 \citet{2011MNRAS.412.1237C} & RAVE DR3 & 13.4 $\pm$ 0.4 \\
 \citet{2010MNRAS.408.1788B} & Masers & 11.0 $\pm$ 1.7\\
Breddels et al. (2010) & RAVE DR2 &  20.4 $\pm$ 0.5 \\
\citet{2007ARep...51..372B} & F $\&$ G stars & 6.2 $\pm$ 2.2 \\
Piskunov et al. (2006) & Open clusters & 11.9 $\pm$ 0.7 \\
\citet{1998MNRAS.298..387D} & Hipparcos & 5.2 $\pm$ 0.6\\
\citet{1965gast.book...61D} & non-MS stars & 12.0 \\
\hline
\end{tabular}
\end{center}
\label{tab:V_ref}
\end{table*}%

We make use of the Str\"omberg relation to estimate the component in the direction of Galactic rotation, V$_{\sun}$, of the Sun's velocity with respect to the Local Standard of Rest (LSR) using a very local and pure thin disc sample. Theoretically, the mean velocity of a group of stars lags behind the circular velocity by an amount that is proportional to its radial velocity dispersion, $\sigma_{\rm U}^{2}$ in our case \citep{1946ApJ...104...12S}. Note that, as pointed out in \citet{2010MNRAS.403.1829S}, applying the Str\"omberg relation to samples of stars binned in colour can be problematic. The linear approximation breaks down due to the metallicity gradient in the disc and hence the extrapolation to zero velocity dispersion becomes invalid. Although this affects some of the works presented in Table~\ref{tab:V_ref} (for example \citealt{1998MNRAS.298..387D}), our result should be robust as the stars are grouped according to chemistry rather than colour.

Following \citet{2013A&A...557A..92G} and using the standard application of the non-linear equation for the asymmetric drift, V$_{a}$, where the quadratic terms $\delta$ V$^{2}$ and V$_{\sun}$$^{2}$ are neglected, we can write the linear Str\"omberg relation as follow,

\begin{equation}
V_{a} = \delta V - V_{\sun} = \frac{\sigma_{\rm U}^{2}}{\it k}
\end{equation}

\noindent where --$\delta$V = $<$V$>$$_{i}$, the mean rotational velocity for the $ith$ chemical group. The slope $k$ depends on the radial scalelength and shape and orientation of the velocity dispersion ellipsoid of the subpopulations.  

 
Using only the thin disc chemical groups (black dots in Fig.~\ref{fig:stromberg_relation}) and fitting a straight line taking into account the intrinsic dispersion in $<$V$>$$_{i}$ uncertainties we determine a solar space velocity, V$_{\sun}$, with respect to the LSR of V$_{\sun}$ = 13.9 $\pm$ 3.4 km s$^{-1}$. This result is in good agreement with several recent studies (see Tab.~\ref{tab:V_ref}) using different stellar populations, though several studies also have found a value close to 5 km s$^{-1}$ \citep{1998MNRAS.298..387D, 2007ARep...51..372B, 2007A&A...474..653V}. One reason for these discrepancies may arise from local kinematical substructure or any systematic streaming motion in the Sun's vicinity \citep{2013MNRAS.436..101W}. A local spiral arm density wave can lead to kinematical fluctuations \citep{2012MNRAS.425.2335S}, and systematic streaming velocities may also exist in the local Galactic disk due to perturbations from the bar \citep{2010MNRAS.407.2122M,2014A&A...563A..60A}. These could cause deviations of the zero-dispersion LSR orbit from the average circular velocity at $R_{0}$ \citep{2014ApJ...793...51S,2015ApJ...800...83B}. We find that $\delta$V for most of the thin disc range from 0 to 20 km s$^{-1}$ (excluding one thin disc group with $\delta$V $\sim$ 40 km s$^{-1}$), while for the thick disc groups we have 40 $<$ $\delta$V $<$ 75 km s$^{-1}$, in good agreement with the mean rotational lag of 51 $\pm$ 5 km s$^{-1}$ for the thick disc reported in \citet{2003A&A...398..141S} using around 900 red clump giants, and also in good agreement with the rotational lag range reported more recently in \citet{2016A&A...596A..98A} for the thick disc using \emph{Gaia} DR1 and SDSS APOGEE.  

\begin{table*}
\caption{Number of stars in each chemical group (N) together with the average abundances values ($<$[Fe/H]$>$, $<$[$\alpha$/Fe]$>$), standard deviation ratios, vertex deviation, tilt of the velocity ellipsoid and the velocity anisotropy parameter for the twelve chemical groups.}
\begin{center}
\begin{tabular}{ccccccccc}
\hline
\hline
& N & $<$[Fe/H]$>$ (dex) & $<$[$\alpha$/Fe]$>$ (dex) & $\sigma_{\rm V}/\sigma_{\rm U}$ & $\sigma_{\rm W}/\sigma_{\rm U}$ & l$_{\rm uv}$ ($^\circ$) & l$_{\rm uw}$ ($^\circ$) & $\beta$ \\
\hline
Thin disc & 41 & +0.07 $\pm$ 0.07 & +0.01 $\pm$ 0.01 & 0.85 $\pm$ 0.11 & 0.62 $\pm$ 0.13 & +35.88 $\pm$ 7.83 &  --10.92 $\pm$ 7.60 &  0.45 $\pm$ 0.15 	\\
               & 25 & --0.07 $\pm$ 0.06 & +0.03 $\pm$ 0.02 & 0.59 $\pm$ 0.12 &  0.57 $\pm$ 0.11 & +14.23 $\pm$ 4.77 &  +7.12 $\pm$ 2.36 & 0.66 $\pm$ 0.11 	\\
	      & 21 & +0.23 $\pm$ 0.04 & +0.01 $\pm$ 0.02  &  0.90 $\pm$ 0.19 & 0.67 $\pm$ 0.14 & +38.51 $\pm$ 14.52 &  +14.49 $\pm$ 15.46 & 0.35 $\pm$ 0.22 \\
	      & 19 & --0.02 $\pm$ 0.06 & +0.04 $\pm$ 0.01  & 0.65 $\pm$ 0.15 & 0.56 $\pm$ 0.12 & +11.21 $\pm$ 8.69 & --2.40 $\pm$ 3.16 & 0.62 $\pm$ 0.14 \\
	      & 16 & --0.52 $\pm$ 0.05 & +0.10 $\pm$ 0.02  &  0.61 $\pm$ 0.15 & 0.77 $\pm$ 0.19 & --5.25 $\pm$ 8.82 & +5.40 $\pm$ 7.30  &  0.50 $\pm$  0.19 \\
	      & 15 & --0.34 $\pm$ 0.04 & +0.09 $\pm$ 0.02  & 0.59 $\pm$ 0.15 & 0.67 $\pm$ 0.17  &  --0.75 $\pm$ 3.89 & --6.69 $\pm$ 7.31 &  0.58 $\pm$ 0.17 \\
	      & 15 &  +0.33 $\pm$ 0.04 & --0.01 $\pm$ 0.02 & 1.07 $\pm$ 0.27 & 1.00 $\pm$ 0.25 & --38.95 $\pm$ 12.14 & --43.68 $\pm$ 12.43 &  --0.11 $\pm$ 0.45 \\
	      \hline
Thick disc & 30 &  --0.29 $\pm$ 0.05 & +0.18 $\pm$ 0.02 & 0.67 $\pm$ 0.12 & 0.80 $\pm$ 0.14 & --16.61 $\pm$ 6.44 &  --6.80 $\pm$  10.12  & 0.44 $\pm$ 0.16 \\
	         & 24 & --0.43 $\pm$ 0.05 & +0.21 $\pm$ 0.02  & 0.54 $\pm$ 0.11 & 0.70 $\pm$ 0.14 & --10.97 $\pm$ 5.27  & --15.16 $\pm$ 8.61 & 0.60 $\pm$ 0.13 \\
	         & 21 & --0.64 $\pm$ 0.07 & +0.26 $\pm$ 0.02 & 0.86 $\pm$ 0.18 & 0.67 $\pm$  0.14  & --7.56 $\pm$ 10.26 & --18.87 $\pm$ 8.20  & 0.39 $\pm$ 0.21 \\
	         & 17 & --0.81 $\pm$ 0.05 & +0.29 $\pm$ 0.01  & 0.69 $\pm$ 0.16 & 0.61 $\pm$ 0.14 & --18.77 $\pm$ 8.76 & +1.26 $\pm$ 1.98  & 0.56 $\pm$ 0.16 \\
	         & 15 & --0.18 $\pm$ 0.06 & +0.13 $\pm$ 0.02 &  0.73 $\pm$ 0.18 & 0.52 $\pm$ 0.13 & +7.55 $\pm$ 10.06 & --6.04 $\pm$ 6.28 & 0.59 $\pm$ 0.16 \\
\hline
\end{tabular}
\end{center}
\label{tab:Fe_AL_beta}
\end{table*}%

\subsection{The vertex deviation and the tilt of the velocity ellipsoid}
\label{vertex}

We can calculate the tilt angles of the mono-abundances groups velocity ellipsoid using Equation~\ref{eq:tilt}. In Figure~\ref{fig:fe_al_vertex} we show the vertex deviation (l$_{uv}$) in the $U$-$V$ plane for the chemical groups as a function of $<$[Fe/H]$>$ and $<$[$\alpha$/Fe]$>$. For the error propagation we use a Monte Carlo technique that takes into account the uncertainties in the velocity dispersion. We generate a distribution  of 10,000  test particles around each input  value of $\sigma_{ii}$ and $\sigma_{ij}$ assuming Gaussian errors  with standard  deviations given  by the  formal errors  in the measurements, and then we calculate the resulting l$_{uv}$ and their corresponding 1$\sigma$ associated uncertainties. For the thin disc we find a strong correlation between the vertex deviation and abundance (black dots in Fig.~\ref{fig:fe_al_vertex}), except for the most metal-rich group: l$_{uv}$ range from +40$^\circ$ for the metal-rich groups to -5$^\circ$ for the most metal-poor and $\alpha$-enhanced thin disc chemical group. Curiously, for the most metal-rich group, l$_{uv}$ $\sim$ -40$^\circ$. The vertex deviation for most of the thick disc mono-abundances groups ($<$[$\alpha$/Fe]$>$ $>$ 0.15 dex) is independent of the abundances (red dots in Fig.~\ref{fig:fe_al_vertex}). For most of the thick disc groups, l$_{uv}$ $\sim$ -15$^\circ$. There is only one thick disc group ($<$[Fe/H]$>$ = -0.18, $<$[$\alpha$/Fe]$>$ = +0.13 dex) that has l$_{uv}$ $>$ 0.

Figure~\ref{fig:fe_al_tilt} shows the tilt angle (l$_{uw}$) in the $U$-$W$ plane for the groups. A positive value means the velocity ellipsoid is tilted towards the Galactic plane toward the Galactic centre. For the thin disc, l$_{uw}$ ranges from -10 to +15$^\circ$, and there is no evident relation between l$_{uw}$ and mean abundance. For the most metal-rich group the tilt angle is around -45$^\circ$. However, for the thick disc groups, we find a weak relation for l$_{uw}$ as a function of iron abundance and $\alpha$-element abundances. The tilt angle decreases from -5 to -20$^\circ$ when [Fe/H] decreases and [$\alpha$/Fe] increases. The most metal-poor and $\alpha$-enhanced group does not follow the common trend for the thick disc, and has a value of l$_{uw}$ $\sim$ +1 $\pm$ 2$^\circ$. We summarize these results in Table~\ref{tab:Fe_AL_beta}.


\begin{figure*}
  \centering  
  \includegraphics[width=2.1\columnwidth]{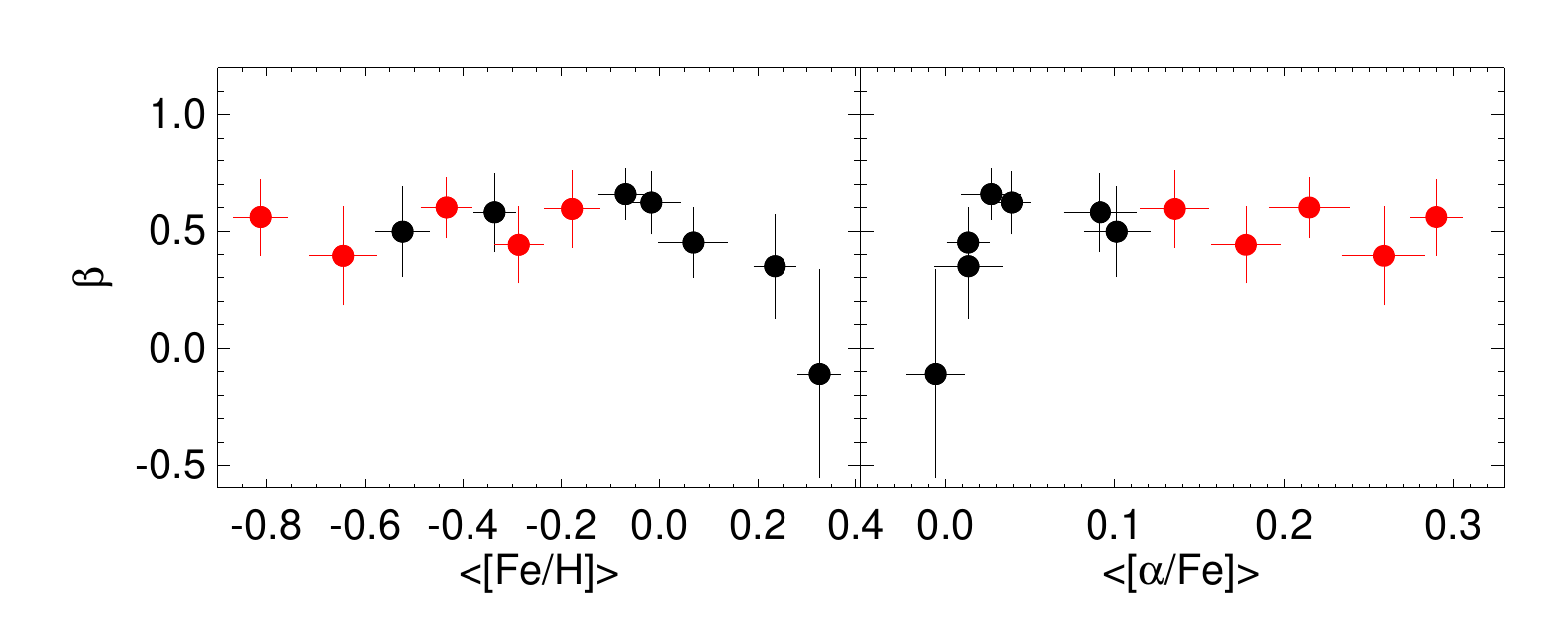}
   \caption{The velocity anisotropy parameter, $\beta$, as a function of $<$[Fe/H]$>$ and $<$[$\alpha$/Fe]$>$ for the thin (black) and thick (red) disc chemical groups studied in this work.}
  \label{fig:fe_al_beta}
\end{figure*}

\subsection{The Galactic velocity anisotropy parameter}

The velocity anisotropy parameter $\beta$, defined as

\begin{equation}
\beta = 1 - \frac{\sigma_{V}^{2} + \sigma_{W}^{2}}{2\sigma_{U}^{2}}
\end{equation}

was introduced by \citet{1980MNRAS.190..873B} to characterize the orbital structure of a system and is extensively used in spherical Jeans equation modeling to estimate the mass distribution of galactic systems (e.g., \citealt{2017ApJ...835..193E} and references therein). By definition, $\beta$ = 0 corresponds to an isotropic velocity distribution ($\sigma_{U}$ = $\sigma_{V}$ = $\sigma_{W}$). If radial or circular orbits dominate (radial anisotropy), $\beta$ is positive (0 $<$ $\beta$ $<$ 1) or negative (--$\infty$ $<$ $\beta$ $<$ 0). 

In Figure~\ref{fig:fe_al_beta} we present $\beta$ as a function of $<$[Fe/H]$>$ and $<$[$\alpha$/Fe]$>$ for the 12 chemical groups analyzed in this work. We find that $\beta$ is independent of the abundances of the groups. For the thin and thick disc chemical groups, our results show $<$$\beta$$>$ $\sim$ 0.5, which suggests that the bulk of the stars in the immediate solar neighborhood are dynamically relaxed, where radial and circular orbits dominate independently of [Fe/H] and [$\alpha$/Fe]. We find for the most metal-rich group ($<$[Fe/H]$>$ = +0.3 dex) a $\beta$ = --0.1, corresponding to an isotropic velocity distribution; however the error bar is large for this group (see Fig.~\ref{fig:fe_al_beta} and Tab.~\ref{tab:Fe_AL_beta}).  

\section{Summary and Discussion}
\label{summary}

We have used the twelve most populated chemical groups identified in \citet{2013MNRAS.428.2321M,2014MNRAS.438.2753M}, where very precise (uncertainties in [X/Fe] less than 0.05 dex) individual stellar abundances from \citet{2014A&A...562A..71B}, parallaxes and proper motions from \emph{Gaia} DR1 \citep{2016A&A...595A...2G} together with the radial velocities from CORAVEL \citep{2004A&A...418..989N} were employed to explored the velocity ellipsoid and its tilt, and the vertex deviation, and velocity anisotropy as a function of stellar iron abundances and $\alpha$-elements. Typical uncertainties for the stellar space velocities are just 0.4 km s$^{-1}$. As determined from the iron abundances with respect to [$\alpha$/Fe], our survey includes seven chemical groups associated with the thin Galactic disc, and five associated with the thick disc. 

We use the chemical groups to examine the correlations between Galactic rotational velocity and mean abundance of the groups. \citet{2011ApJ...738..187L}, using SDSS G-dwarfs and more recently \citet{2016A&A...596A..98A}, combining APOGEE data with \emph{Gaia} DR1, reported a statistical relationship showing opposite gradients signs for the thin and thick disc in the rotational velocity as a function of [Fe/H]. A possible explanation for the reverse gradient of rotational velocity is due to inwards radial migration of relatively metal-poor stars with low velocity dispersion and high angular momentum from the outer disc, and more metal-rich stars from the inner disc into the solar vicinity. The angular momentum exchange was only partial, hence their velocity dispersion is still low despite the migration \citep{2011ApJ...738..187L}. In Section~\ref{rot_vel} we look at the $<$V$>$-[Fe/H]-[$\alpha$/Fe] tends for the chemical groups (see Fig.~\ref{fig:V_Fe}). Using all the groups in the thin disc we are able to reproduce the results found in \citet{2011ApJ...738..187L}, and \citet{2016A&A...596A..98A}. However we find no correlation between $<$V$>$ and [$\alpha$/Fe] for the thin disc. Moreover in the $<$V$>$ - [Fe/H] diagram there is one thin disk group in the thick disk sequence, which lies about 20 km s$^{-1}$ lower than the thin disk sequence at its [Fe/H]. How does this chemical group fit into the picture described above? We present an alternative scenario in the bottom-panels of Fig.~\ref{fig:V_Fe}. For the best linear fit in the thin disk, we excluded the most metal-poor groups ---they clearly show a different trend with respect to most of the thin disk groups in the $<$V$>$-[Fe/H]-[$\alpha$/Fe] diagram---. In this picture the thin disc shows the asymmetric drift as a function of both abundances, [Fe/H] and [$\alpha$/Fe], where more metal-poor stars are lagging further behind. More mono-abundance chemical groups with precise kinematics are needed to address this problem specially groups in the metallicity range between -0.1 and -0.3 dex.  
    
We find average velocity dispersion values for the three space velocity components for the thin and thick disc of ($\sigma_{\rm U}$,$\sigma_{\rm V}$,$\sigma_{\rm W}$)$_{\rm thin}$ = (33 $\pm$ 4, 28 $\pm$ 2, 23 $\pm$ 2) and ($\sigma_{\rm U}$,$\sigma_{\rm V}$,$\sigma_{\rm W}$)$_{\rm thick}$ = (57 $\pm$ 6, 38 $\pm$ 5, 37 $\pm$ 4) km s$^{-1}$. The results found in this exercise for the mean velocity dispersion are similar within the errors to those from previous studies, e.g. \citet{2001ASPC..230...87Q}, \citet{2008A&A...480...91S}, \citet{2012A&A...547A..70P} and \citet{2013A&A...554A..44A}. The very metal-poor and $\alpha$-enhancement groups are typically hotter than their more metal-rich and low $\alpha$-enhancement counterparts, though interestingly, not all the chemically selected thick disc chemical groups are hotter than the thin disc ones (see Fig.~\ref{fig:al_fe_disper}). For the $\sigma_{\rm V}$ component there is a clear plateau for the thin disc chemical groups with respect to the iron abundances and [$\alpha$/Fe]. For the most metal-poor (and high-[$\alpha$/Fe]) thick disc groups there is an abrupt increase of $\sigma_{\rm V}$. 

We also present a characterization of the velocity ellipsoid as a function of stellar iron abundance and $\alpha$-element enrichment. The ratio between the semi-axes of the velocity ellipsoid, $\sigma_{\rm V}$/$\sigma_{\rm U}$ and $\sigma_{\rm W}$/$\sigma_{\rm U}$ are related to the Oort constants and the scattering process responsible for the dynamical heating of the Milky Way disc, respectively. From our study we have found that the mean value of $<\sigma_{\rm V}/\sigma_{\rm U}>_{\rm thin}$ is 0.70 $\pm$ 0.13, while $<\sigma_{\rm V}/\sigma_{\rm U}>_{\rm thick}$ is 0.64 $\pm$ 0.08, whereas the result for $<\sigma_{\rm W}/\sigma_{\rm U}>_{\rm thin}$ is 0.67 $\pm$ 0.11, and $<\sigma_{\rm W}/\sigma_{\rm U}>_{\rm thick}$ is 0.66 $\pm$ 0.11. Thus, we do not find different ratios for the thin and thick disc groups. Our values are slightly higher than the results from \citet{1998MNRAS.298..387D}, where they found $\sigma_{\rm V}$/$\sigma_{\rm U}$ = 0.6 and $\sigma_{\rm W}$/$\sigma_{\rm U}$ = 0.5 using \emph{Hipparcos} data. But, using the RAVE survey, \citet{2008A&A...480..753V} found $\sigma_{\rm W}$/$\sigma_{\rm U}$ = 0.9, while \citet{2003A&A...398..141S} found $\sigma_{\rm W}$/$\sigma_{\rm U}$ = 0.6 using only a data-set of red giants, and \citet{2006A&A...451..125V} reported a value of 0.5 for the ratio where stellar populations toward the North Galactic Pole were employed. \citet{2012ApJ...747..101M} showed a mean value of $\sigma_{\rm V}$/$\sigma_{\rm U}$ = 0.73 $\pm$ 0.05 and $\sigma_{\rm W}$/$\sigma_{\rm U}$ = 0.48 $\pm$ 0.06, respectively. \citet{2012ApJ...746..181S} found a range of values for $\sigma_{\rm V}$/$\sigma_{\rm U}$, from 0.7 to 0.9 and from 0.55 to 0.85 for $\sigma_{\rm W}$/$\sigma_{\rm U}$ as a function of $z$ using SDSS data. Simulations of the formation of the thick disc through heating via accretion predict a wide range of values from 0.4 to 0.9 for $\sigma_{\rm W}$/$\sigma_{\rm U}$, depending on satellite mass ratio and orbital inclination \citep{2010ApJ...718..314V}. The ratio $\sigma_{\rm V}$/$\sigma_{\rm U}$ is predicted to lie around 0.5 and 0.6 depending of the shape of the rotation curve \citep{1991dodg.conf...71K}. Interestingly, the super-solar metallicity groups show larger $\sigma_{\rm V}$/$\sigma_{\rm U}$ ratios than the more metal-poor thin disc groups, where these values go from 0.8 to 1.1, but where the uncertainties are also larger. In the $\sigma_{\rm W}$/$\sigma_{\rm U}$ ratio, the most metal-rich group clearly has a larger derived value of 1.0 $\pm$ 0.2 compared with the other groups (see Fig.~\ref{fig:al_fe_disper} and Tab.~\ref{tab:Fe_AL_beta}). Finally, for the thin disc groups there is a trend between $\sigma_{\rm V}$/$\sigma_{\rm U}$ and $<$[Fe/H]$>$ and $<$[$\alpha$/Fe]$>$ ranging from 1.1 to 0.6, except for the groups with $<$[Fe/H]$>$ $<$ --0.2 and $<$[$\alpha$/Fe]$>$ $\sim$ 0.05 dex, where $\sigma_{\rm V}$/$\sigma_{\rm U}$ $\sim$ 0.6. For the thick disc groups we do not see any clear trend between $\sigma_{\rm V}$/$\sigma_{\rm U}$ as a function of iron-abundance and $\alpha$-element abundances. We find $\sigma_{\rm W}$/$\sigma_{\rm U}$ nearly independent of groups abundance.

Using only the thin disc chemical groups and making use of the Str\"omberg relation, we determined the expected velocity component in the direction of Galactic rotation of the Sun's velocity with respect to the Local Standard of Rest (LSR). Our result, V$_{\sun}$ = 13.9 $\pm$ 3.4 km s$^{-1}$, is in good agreement with several recent studies (see Tab.~\ref{tab:V_ref}).

We also explore the orientation of the velocity ellipsoid. The orientation of a dynamically relaxed population is related to the Galactic potential shape, and the vertex deviation can indicate the presence of non-axisymmetric structures in the disc. For the thin disc we find a trend between the vertex deviation and abundances (except for the most metal-rich group), ranging from +40$^\circ$ for the metal-rich groups to -5$^\circ$ for the metal-poor thin disc. The vertex deviation for the thick disc mono-abundances groups is independent of abundance, with a value of l$_{uv}$ $\sim$ -15$^\circ$. Substructure observed in the velocity space might be responsible for the vertex deviation, however it is not totally clear that moving groups are entirely responsible for the non-zero values of l$_{\rm uv}$. Additional causes of the vertex deviation could be the non-axisymmetric component of the Galactic potential and the spiral field \citep{2005A&A...430..165F,2008MNRAS.391..793S}. Why do for most of the thick disc groups we find a nearly constant vertex deviation? Significant overdensities in the velocity distributions are reported beyond the solar circle; for example, \citet{2012MNRAS.426L...1A} found that local kinematical groups are large-scale features, surviving at least up to $\sim$ 1 kpc from the Sun at different heights from the Galactic plane. Some of these trends are consistent with dynamical models of the effects of the bar and the spiral arms; however, in this study we find a nearly constant vertex deviation for the Galactic thick disc despite the fact a significant variation may be expected from the non-axisymmetric components associated with the thick disc.  

 The tilt angle (l$_{uw}$) in the $U$-$W$ plane for the thin disc groups range from -10 to +15$^\circ$, and there is no evident correlation between l$_{uw}$ and the mean abundances. Furthermore we find a weak trend of l$_{uw}$ as a function of iron abundance and $\alpha$-element abundances for most of the groups in the thick disc, where the tilt angle decreases from -5 to -20$^\circ$ when [Fe/H] decreases and [$\alpha$/Fe] increases. Also, the tilt of the most metal-rich group seems to deviate from the common trend, as observed for the vertex deviation. The tilt of the velocity ellipsoid is a good indicator of the buckling instability of a stellar bar in a disk galaxy \citep{2013ApJ...764..123S}. This may suggest that this group could be an overlap of spiral and bar resonances in the Milky way disc \citep{2010ApJ...722..112M}. The velocity anisotropy parameter, $\beta$, is independent of the chemical group abundances and its value is nearly constant for both discs, with a mean value of $<$$\beta$$>$ $\sim$ 0.5, which suggests that the disc is dynamically relaxed. The most metal-rich group has $\beta$ $\sim$ 0, corresponding to an isotropic velocity distribution, but the uncertainties here are large.


In this work we use a small sample with very precise stellar abundances and kinematics, however the APOGEE  and APOGEE-2 surveys \citep{2017AJ....154...94M} have accurate individual abundances for more than 300,000 stars in both celestial hemisphere covering a much larger volume than the present study. The GALAH survey \citep{2015MNRAS.449.2604D} has already observed more than half million stars in the South with high-resolution and high signal-to-noise ratio. Combined with astrometry from \emph{Gaia}, these samples are very promising data-sets to explore in great detail though mono-abundances groups via chemical tagging the kinematical properties of the Galactic disc. Such work is underway, and will be the subject of future contributions. 


\section*{Acknowledgments}

The authors thank the referee for comments that helped improve the paper. BA thanks Hanna Lewis, Chris Hayes, Nick Troup and Robert Wilson (University of Virginia) for lively discussions on the manuscript. BA and SRM acknowledge support from National Science Foundation grant AST-1616636.

\appendix

\bsp



\label{lastpage}

\end{document}